\documentclass[twocolumn]{aastex631}

\newcommand{\e}{${\rm e^{-}}$}

\newcommand{\lya}{Ly$\alpha$}

\usepackage{booktabs}
\usepackage{hyperref}
\usepackage{amsmath}
\usepackage[mathlines]{lineno}
\usepackage{graphicx}
\usepackage{graphbox}
\usepackage{enumitem}
\usepackage{bm}
\usepackage{color}
\usepackage[normalem]{ulem}
\usepackage{natbib}
\defcitealias{Liang+2021}{L21}

\shorttitle{Cosmic Himalayas: Discrepancies between Galaxies, Quasars, and IGM}
\shortauthors{Liang et al.}
\graphicspath{{./}{figures/}}
\begin{document}

\title{
\textit{Cosmic Himalayas}: The Highest Quasar Density Peak Identified in a 10,000 deg$^2$ Sky \\ with Spatial Discrepancies between Galaxies, Quasars, 
and IGM {\sc Hi}

% Monsters Dance on the Stage with Dual Tracks (MONDSTADT). I. The First Glimpse via IGM Tomography Built by SDSS on the Most Concentrated Quasars at Cosmic Noon.
}

\correspondingauthor{Yongming Liang}
\email{ymliang@icrr.u-tokyo.ac.jp}

\author[0000-0002-2725-302X]{Yongming Liang}
\affil{Institute for Cosmic Ray Research, The University of Tokyo, 5-1-5 Kashiwanoha, Kashiwa, Chiba 277-8582, Japan}
\affil{National Astronomical Observatory of Japan, 2-21-1 Osawa, Mitaka, Tokyo 181-8588, Japan}

\author[0000-0002-1049-6658]{Masami Ouchi}
\affil{Institute for Cosmic Ray Research, The University of Tokyo, 5-1-5 Kashiwanoha, Kashiwa, Chiba 277-8582, Japan}
\affil{National Astronomical Observatory of Japan, 2-21-1 Osawa, Mitaka, Tokyo 181-8588, Japan}
\affil{Department of Astronomical Science, SOKENDAI (The Graduate University for Advanced Studies), Mitaka, Tokyo 181-8588, Japan}
\affil{Kavli IPMU (WPI), The University of Tokyo, 5-1-5 Kashiwanoha, Kashiwa, Chiba 277-8583, Japan}

\author[0000-0002-1199-6523]{Dongsheng Sun}
\affil{Institute for Cosmic Ray Research, The University of Tokyo, 5-1-5 Kashiwanoha, Kashiwa, Chiba 277-8582, Japan}

\author[0000-0003-3954-4219]{Nobunari Kashikawa}
\affil{Department of Astronomy, School of Science, The University of Tokyo, 7-3-1 Hongo, Bunkyo-ku, Tokyo, 113-0033, Japan}
\affil{Research Center for the Early Universe, The University of Tokyo, 7-3-1 Hongo, Bunkyo-ku, Tokyo 113-0033, Japan}

\author[0000-0001-8467-6478]{Zheng Cai}
\affil{Department of Astronomy, Tsinghua University, Beijing 100084, China}

\author[0000-0001-5804-1428]{Sebastiano Cantalupo}
\affil{Department of Physics, University of Milan Bicocca, Piazza della Scienza 3, I-20126 Milano, Italy}

\author[0000-0001-7457-8487]{Kentaro Nagamine}
\affil{Department of Earth and Space Science, Osaka University, 1-1 Machikaneyama, Toyonaka, Osaka 560-0043, Japan}
\affil{Kavli IPMU (WPI), The University of Tokyo, 5-1-5 Kashiwanoha, Kashiwa, Chiba 277-8583, Japan}
\affil{Department of Physics and Astronomy, University of Nevada, Las Vegas, 4505 S. Maryland Pkwy, Las Vegas, NV 89154-4002, USA}

\author[0000-0002-1319-3433]{Hidenobu Yajima}
\affil{Center for Computational Sciences, University of Tsukuba, 1-1-1 Tennodai, Tsukuba, Ibaraki 305-8577, Japan}

\author[0000-0001-6503-8315]{Takanobu Kirihara}
\affil{Kitami Institute of Technology, 165 Koen-cho, Kitami, Hokkaido 090-8507, Japan}
\affil{Center for Computational Sciences, University of Tsukuba, 1-1-1 Tennodai, Tsukuba, Ibaraki 305-8577, Japan}

\author[0000-0003-2273-9415]{Haibin Zhang}
\affil{National Astronomical Observatory of Japan, 2-21-1 Osawa, Mitaka, Tokyo 181-8588, Japan}

\author[0000-0001-6251-649X]{Mingyu Li}
\affil{Department of Astronomy, Tsinghua University, Beijing 100084, China}

\author[0000-0003-4442-2750]{Rhythm Shimakawa}
\affil{Waseda Institute for Advanced Study (WIAS), Waseda University, 1-21-1 Nishi-Waseda, Shinjuku, Tokyo 169-0051, Japan}
\affil{Center for Data Science, Waseda University, 1-6-1 Nishi-Waseda, Shinjuku, Tokyo 169-0051, Japan}

\author[0000-0003-3310-0131]{Xiaohui Fan}
\affil{Steward Observatory, University of Arizona, 933 N. Cherry Avenue, Tucson, AZ 85721, USA}

\author[0000-0002-9453-0381]{Kei Ito}
\affil{Cosmic Dawn Center (DAWN), Denmark}
\affil{DTU Space, Technical University of Denmark, Elektrovej 327, DK2800 Kgs. Lyngby, Denmark}
\affil{Department of Astronomy, School of Science, The University of Tokyo, 7-3-1 Hongo, Bunkyo-ku, Tokyo, 113-0033, Japan}

\author[0000-0002-5011-5178]{Masayuki Tanaka}
\affil{Department of Astronomical Science, SOKENDAI (The Graduate University for Advanced Studies), Mitaka, Tokyo 181-8588, Japan}
\affil{National Astronomical Observatory of Japan, 2-21-1 Osawa, Mitaka, Tokyo 181-8588, Japan}

\author[0000-0002-6047-430X]{Yuichi Harikane}
\affil{Institute for Cosmic Ray Research, The University of Tokyo, 5-1-5 Kashiwanoha, Kashiwa, Chiba 277-8582, Japan}

\author[0000-0002-7738-6875]{J. Xavier Prochaska}
\affil{UCO/Lick Observatory, University of California, 1156 High Street, Santa Cruz, CA 95064, USA}

\author[0000-0002-8863-888X]{Andrea Travascio}
\affil{Department of Physics, University of Milan Bicocca, Piazza della Scienza 3, I-20126 Milano, Italy}

\author[0000-0002-9593-8274]{Weichen Wang}
\affil{Department of Physics, University of Milan Bicocca, Piazza della Scienza 3, I-20126 Milano, Italy}

\author[0000-0001-5060-1398]{Martin Elvis}
\affil{Harvard-Smithsonian Center for Astrophysics, 60 Garden St., Cambridge, MA 02138, USA}

\author[0000-0002-3554-3318]{Giuseppina Fabbiano}
\affil{Harvard-Smithsonian Center for Astrophysics, 60 Garden St., Cambridge, MA 02138, USA}

\author[0009-0007-0864-7094]{Junya Arita}
\affil{Department of Astronomy, School of Science, The University of Tokyo, 7-3-1 Hongo, Bunkyo-ku, Tokyo, 113-0033, Japan}

\author[0000-0003-2984-6803]{Masafusa Onoue}
\affil{Kavli IPMU (WPI), The University of Tokyo, 5-1-5 Kashiwanoha, Kashiwa, Chiba 277-8583, Japan}

\author[0000-0002-0000-6977]{John D. Silverman}
\affil{Kavli IPMU (WPI), The University of Tokyo, 5-1-5 Kashiwanoha, Kashiwa, Chiba 277-8583, Japan}

\author[0000-0002-4314-5686]{Dong Dong Shi}
\affil{Center for Fundamental Physics, School of Mechanics and Optoelectronic Physics, Anhui University of Science and Technology, Huainan, Anhui 232001, China}

\author[0000-0001-7943-0166]{Fangxia An}
\affil{Yunnan Observatories, Chinese Academy of Sciences, Kunming 650011, People’s Republic of China}

\author[0000-0001-9452-0813]{Takuma Izumi}
\affil{National Astronomical Observatory of Japan, 2-21-1 Osawa, Mitaka, Tokyo 181-8588, Japan}

\author[0000-0002-2597-2231]{Kazuhiro Shimasaku}
\affil{Department of Astronomy, School of Science, The University of Tokyo, 7-3-1 Hongo, Bunkyo-ku, Tokyo, 113-0033, Japan}
\affil{Research Center for the Early Universe, The University of Tokyo, 7-3-1 Hongo, Bunkyo-ku, Tokyo 113-0033, Japan}

\author[0000-0002-0673-0632]{Hisakazu Uchiyama}
\affil{National Astronomical Observatory of Japan, 2-21-1 Osawa, Mitaka, Tokyo 181-8588, Japan}

\author[0000-0002-9888-6895]{Chenghao Zhu}
\affil{Institute for Cosmic Ray Research, The University of Tokyo, 5-1-5 Kashiwanoha, Kashiwa, Chiba 277-8582, Japan}



\begin{abstract}

We report the identification of a quasar overdensity in the BOSSJ0210 field, dubbed \textit{Cosmic Himalayas}, consisting of 11 quasars at $z=2.16-2.20$, the densest overdensity of quasars ($17\sigma$) in the $\sim$10,000 deg$^2$ of the Sloan Digital Sky Survey. We present the spatial distributions of galaxies and quasars and an \textsc{Hi} absorption map of the intergalactic medium (IGM). On the map of 465 galaxies selected from the MAMMOTH-Subaru survey, we find two galaxy density peaks that do not fall on the quasar overdensity but instead exist at the northwest and southeast sides, approximately 25 $h^{-1}$ comoving-Mpc apart from the quasar overdensity. With a spatial resolution of 15 $h^{-1}$ comoving Mpc in projection, we produce a three-dimensional \textsc{Hi} tomography map by the IGM Ly$\alpha$ forest in the spectra of 23 SDSS/eBOSS quasars behind the quasar overdensity. Surprisingly, the quasar overdensity coincides with neither an absorption peak nor a transmission peak of IGM \textsc{Hi} but lies near the border separating opaque and transparent volumes, with the more luminous quasars located in an environment with lesser IGM \textsc{Hi}. Hence remarkably, the overdensity region traced by the 11 quasars, albeit all in coherently active states, has no clear coincidence with peaks of galaxies or \textsc{Hi} absorption densities. Current physical scenarios with mixtures of \textsc{Hi} overdensities and quasar photoionization cannot fully interpret the emergence of \textit{Cosmic Himalayas}, suggesting this peculiar structure is an excellent laboratory to unveil the interplay between galaxies, quasars, and the IGM.
\end{abstract}



\section{Introduction} \label{sec:intro}

Quasars, or quasi-stellar objects (QSOs), represent some of the most luminous and extreme entities in the cosmos, characterized by their stellar-like appearance. This appearance is attributed to the intense luminosity emanating from the active galactic nucleus (AGN), where material actively accretes onto a supermassive black hole (SMBH) at the galactic center \citep[e.g.,][]{Hopkins+2006}.
Such extreme systems are pivotal to the fields of astronomy and cosmology, offering insights into the growth mechanisms of SMBHs in the early universe and aiding in our understanding of AGN feedback to galaxy formation and evolution \citep{Fabian+2012}.

Quasars are recognized as remarkable tracers of the universe's large-scale structure (LSS).
Typically consisting of at least five quasars  
and spanning sizes greater than $100$ $h^{-1}$ comoving-Mpc (cMpc), Large quasar groups (LQGs), 
such as the one first discovered by \citet{Webster+1982}, have predominantly been identified at relatively low redshifts ($z<2$) through both initial surveys and later, more comprehensive redshift surveys like 2dF and SDSS \citep{Clowes+1986, Crampton+1989, Clowes+1991, Komberg+1996, Clowes+2012, Clowes+2013, Park+2015}. 
The presence of these structures poses questions regarding their alignment with predictions of standard LSS formation theories \citep{Clowes+2013, Pilipenko+2013, Nadathur+2013}. 

Beyond redshift $z=2$, in the period known as Cosmic Noon, environments of high density are considered precursors to contemporary galaxy clusters, i.e., protoclusters \citep{Cen+2000, Overzier+2016}. 
Luminous quasars, including those that are radio-loud, have been identified as convenient tracers
for protoclusters, facilitating their detection and examination \citep{Wylezalek+2013, Onoue+2018}, while some opposite suggestions also persist \citep{Kashikawa+2007, 
Uchiyama+2018, Uchiyama+2020}. 
However, the role of fainter AGNs and their influence on the clustering properties of galaxies in these early cosmic structures was initially overlooked. 
Recent studies now indicate that AGNs, irrespective of their luminosity, significantly contribute to the formation and governance of galaxy populations within these dense early universe environments  \citep{Ito+2022, Shimakawa+2023b}.

A promising method to study the distribution of matter at these high redshifts is three-dimensional (3D) intergalactic medium (IGM) tomography. 
By analyzing the Ly$\alpha$ absorption in the spectra of background quasars, researchers can map the three-dimensional distribution of hydrogen gas, tracing the filaments and voids that constitute the LSSs \citep{Lee+2014a, Lee+2016, Lee+2018, Newman+2020, Horowitz+2022, Sun+2023}. 
This technique provides an unparalleled perspective on the interplay between quasars, galaxy formation, and the IGM \citep{Newman+2022, Dong+2023}.

In this study, we report the discovery of \textit{Cosmic Himalayas}, a significant quasar overdensity within the J0210 field at $z \approx 2.16-2.20$.
Eleven quasars are identified within a volume of (40 cMpc)$^3$ from SDSS-IV/eBOSS. 
Previous observations utilized narrowband and $g$-band imaging from the Subaru Telescope's Hyper Suprime-Cam \citep[HSC;][]{Miyazaki+2018} to study the galaxy--IGM correlation in this field \citep[][hereafter \citetalias{Liang+2021}]{Liang+2021}. 
They target strong \lya~absorption in the IGM, known as coherently strong Ly$\alpha$ absorbers (CoSLAs), utilizing Sloan Digital Sky Survey (SDSS)/eBOSS spectra of IGM absorbers to identify massive structures within the Mapping the Most Massive Overdensity Through Hydrogen with Subaru project \citep[MAMMOTH-Subaru;][Y. Liang et al., in prep.]{Cai+2016, Cai+2017b, Liang+2021, Ma+2024, ZhangHB+2024, LiMY+2024}.
Therefore, J0210 also stands out as a region with exceptional coherent Ly$\alpha$ absorption. 

This paper is organized as follows: 
Section \ref{sec:data} introduces the data and observations utilized in this study. 
In Section \ref{sec:qso_lae}, we analyze the spatial distribution of SDSS quasars and \lya~emitters (LAEs), highlighting the identification of the \textit{Cosmic Himalayas} and its significance within the LSS. 
Section \ref{sec:hi_tomography} presents our methodology and results from the 3D IGM \textsc{Hi} tomography.
Finally, Section \ref{sec:discussion} discusses the implications of our findings on 
the association between the clustered quasars and IGM hydrogen, the possible scenarios resolving a peculiar ionized structure, and the potential triggers of \textit{Cosmic Himalayas}. 
We conclude with a summary of our key results in the final section.
Throughout the paper, we employ cosmology with $\Omega_m=0.3$, $\Omega_{\Lambda}=0.7$, and $h=0.7$. 
AB magnitudes are used if there is no specific mention. 

\section{Data} \label{sec:data}

\subsection{SDSS Quasars}

\begin{table*}[hbt!]
\hspace{-25mm}
\begin{tabular}{@{}cccccccccc@{}}
\toprule
ID & RA (J2000) & DEC (J2000) & Redshift & Plate & MJD & Fiber & $i$-Mag (AB) &$L_{\rm bol}/10^{46}$ erg s$^{-1} $ & $\delta_{\rm F}$ \\ 
$[1]$ & $[2]$ & $[3]$ & $[4]$ & $[5]$ & $[6]$ & $[7]$ & $[8]$ & $[9]$ & $[10]$  \\ \midrule
CH-Q01 & 02:09:53.16 & +00:55:11.0 & 2.1891$\pm$0.0002 & 9383 & 58097 & 585 & 18.89$\pm$0.01 & 6.58$\pm$3.29 & 0.131 \\
CH-Q02 & 02:09:53.48 & +00:51:13.5 & 2.1907$\pm$0.0007 & 702 & 52178 & 535 & 19.58$\pm$0.02 & 2.74$\pm$1.37 & 0.160 \\
CH-Q03 & 02:09:21.92 & +00:51:48.9 & 2.1669$\pm$0.0005 & 7833 & 57286 & 122 & 21.01$\pm$0.05 & 2.02$\pm$1.01 & 0.186 \\
CH-Q04 & 02:10:09.69 & +00:39:51.5 & 2.1691$\pm$0.0004 & 7834 & 56979 & 635 & 21.42$\pm$0.10 & 1.01$\pm$0.50 & 0.084 \\
CH-Q05 & 02:10:00.50 & +00:47:27.0 & 2.1748$\pm$0.0003 & 7833 & 57286 & 126 & 20.47$\pm$0.04 & 0.95$\pm$0.48 & 0.123 \\
CH-Q06 & 02:09:59.37 & +01:06:48.7 & 2.1962$\pm$0.0003 & 7833 & 57286 & 134 & 21.10$\pm$0.06 & 0.83$\pm$0.41 & 0.119 \\
CH-Q07 & 02:09:52.55 & +00:43:08.0 & 2.1715$\pm$0.0004 & 7834 & 56979 & 624 & 21.33$\pm$0.08 & 0.79$\pm$0.40 & 0.184 \\
CH-Q08 & 02:10:24.75 & +00:53:57.7 & 2.1693$\pm$0.0009 & 4235 & 55451 & 834 & 21.78$\pm$0.13 & 0.73$\pm$0.36 & 0.081 \\
CH-Q09 & 02:10:22.06 & +00:57:45.0 & 2.2000$\pm$0.0006 & 4235 & 55451 & 832 & 21.69$\pm$0.10 & 0.50$\pm$0.25 & -0.096 \\
CH-Q10 & 02:10:13.03 & +00:56:13.5 & 2.1997$\pm$0.0005 & 7834 & 56979 & 621 & 21.64$\pm$0.11 & 0.46$\pm$0.23 & 0.019 \\
CH-Q11 & 02:10:14.29 & +00:53:13.5 & 2.1921$\pm$0.0008 & 7834 & 56979 & 623 & 21.32$\pm$0.08 & 0.29$\pm$0.15 & 0.017 \\
\bottomrule
\end{tabular}
\caption{
SDSS-IV/eBOSS Quasar Information for the \textit{Cosmic Himalayas}: 
$[1]$ Quasar ID;
$[2]$ Right Ascension (RA), equinox J2000;
$[3]$ Declination (DEC), equinox J2000;
$[4]$ Spectroscopic redshift from both PCA fitting in the SDSS-IV/eBOSS pipeline and visual confirmation; 
$[5]$ SDSS plate ID; 
$[6]$ SDSS Modified Julian Date (MJD); 
$[7]$ SDSS fiber ID;
$[8]$ SDSS $i$-band magnitude;
$[9]$ Bolometric luminosity ($L_{\rm bol}$) derived from $L_{1350}$;
$[10]$ Ly$\alpha$ transmission fluctuation ($\delta_F$) at the quasar's location.}
\label{tab:qso_info}
\end{table*}

\begin{figure*}[hbt!]
    \centering
    \includegraphics[width=0.45\linewidth]{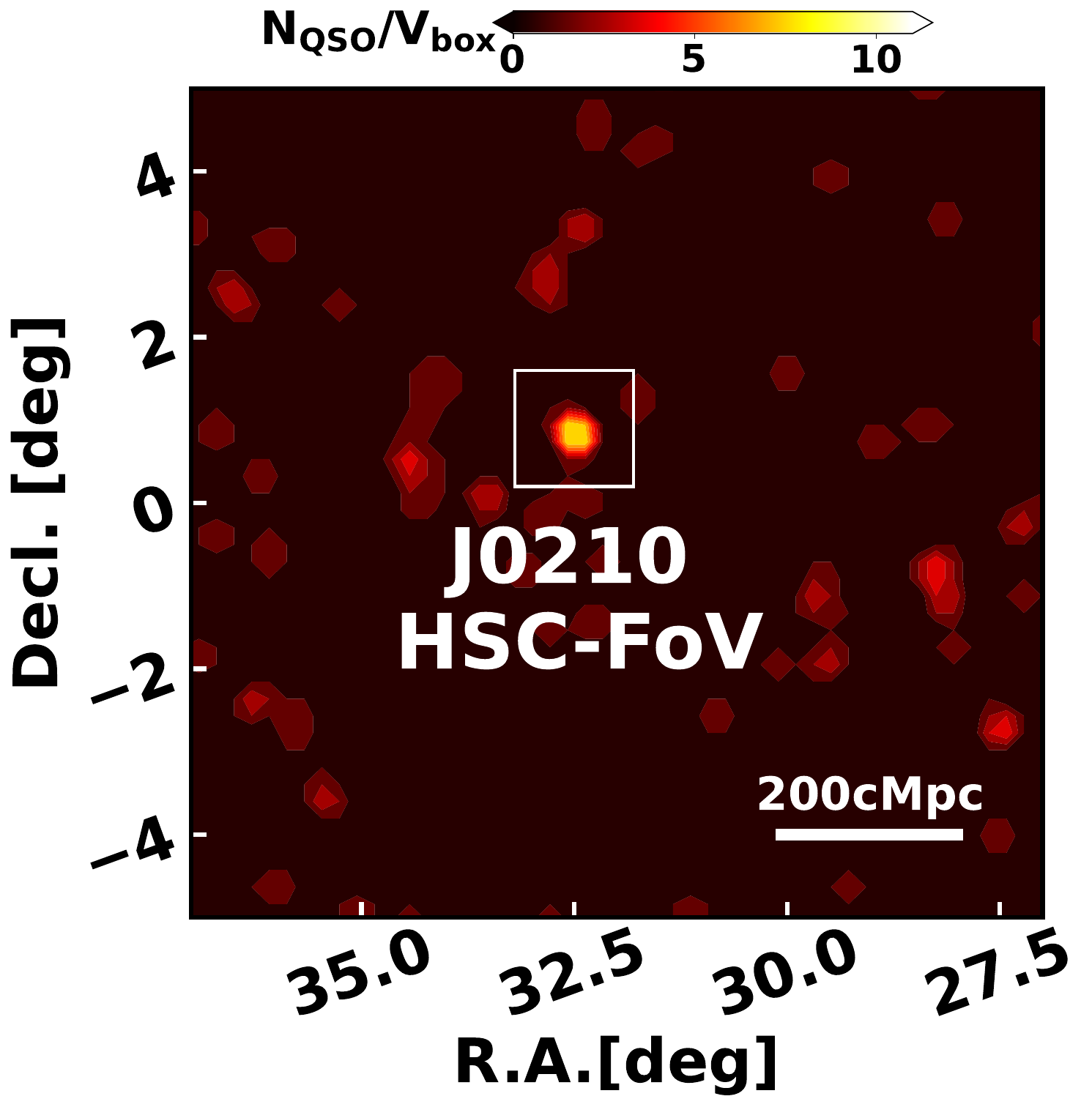}
    \includegraphics[width=0.54\linewidth]{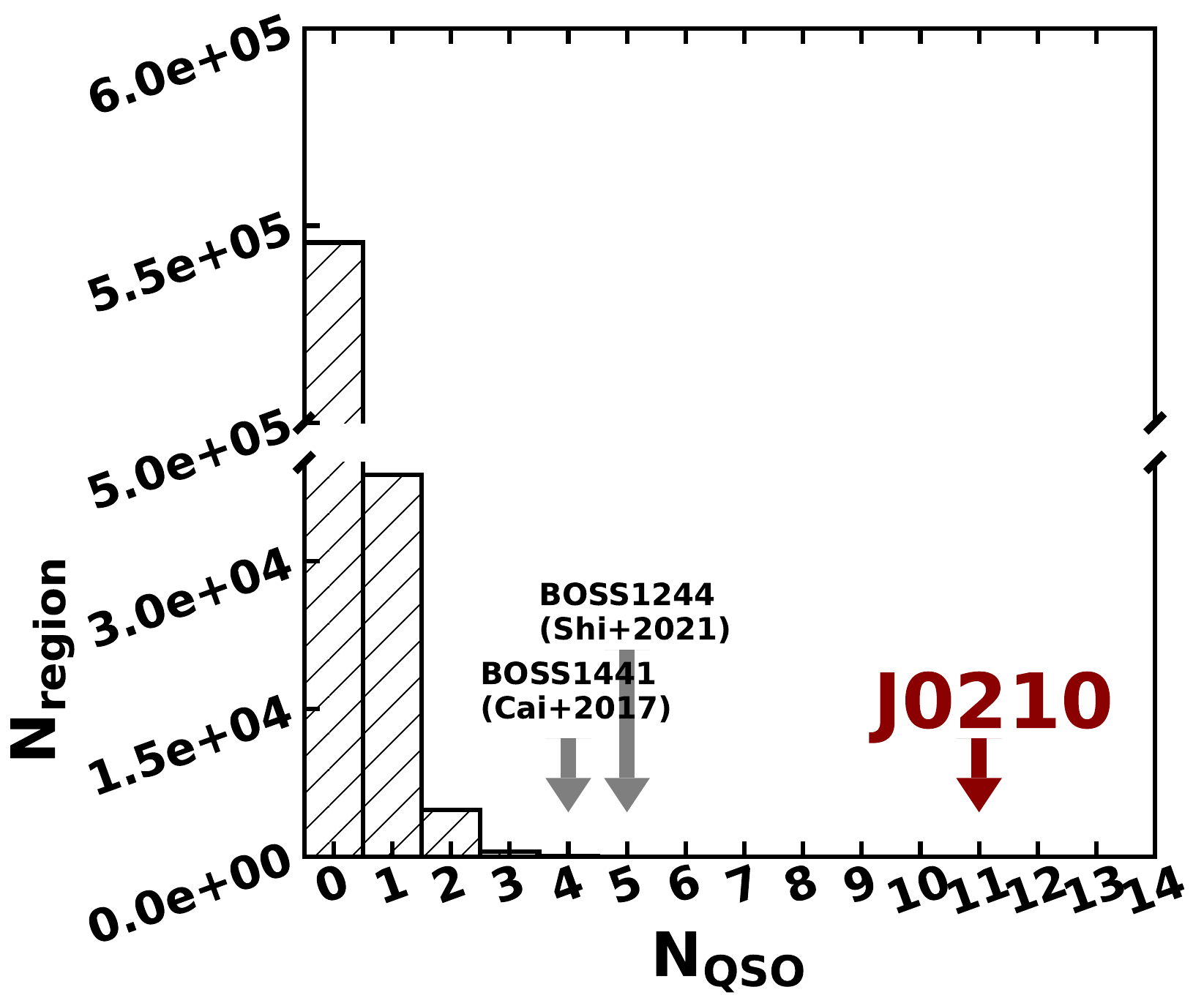}
    \caption{
    {\it Left (a):} Sky distribution of a
    part of SDSS quasars, with whiter shades indicating higher quasar densities.
    {\it Right (b):} Number counts of SDSS quasars within a $\left(40~{\rm cMpc}\right)^3$ volume across the entire 10,000 deg$^2$ survey area with a step size of 20 cMpc along each dimension. 
    The average count is 0.36, with J0210 hosting 11 quasars, marking a density nearly 30 times the mean. For context, other fields with high SDSS quasar densities are highlighted \citep{Cai+2017b, Shi+2021}; even these are
    less than half of the overdensity found for J0210.
    }
    \label{fig:qso_sky}
\end{figure*}

\begin{figure}[hbt!]
    \centering
    \includegraphics[width=\linewidth]{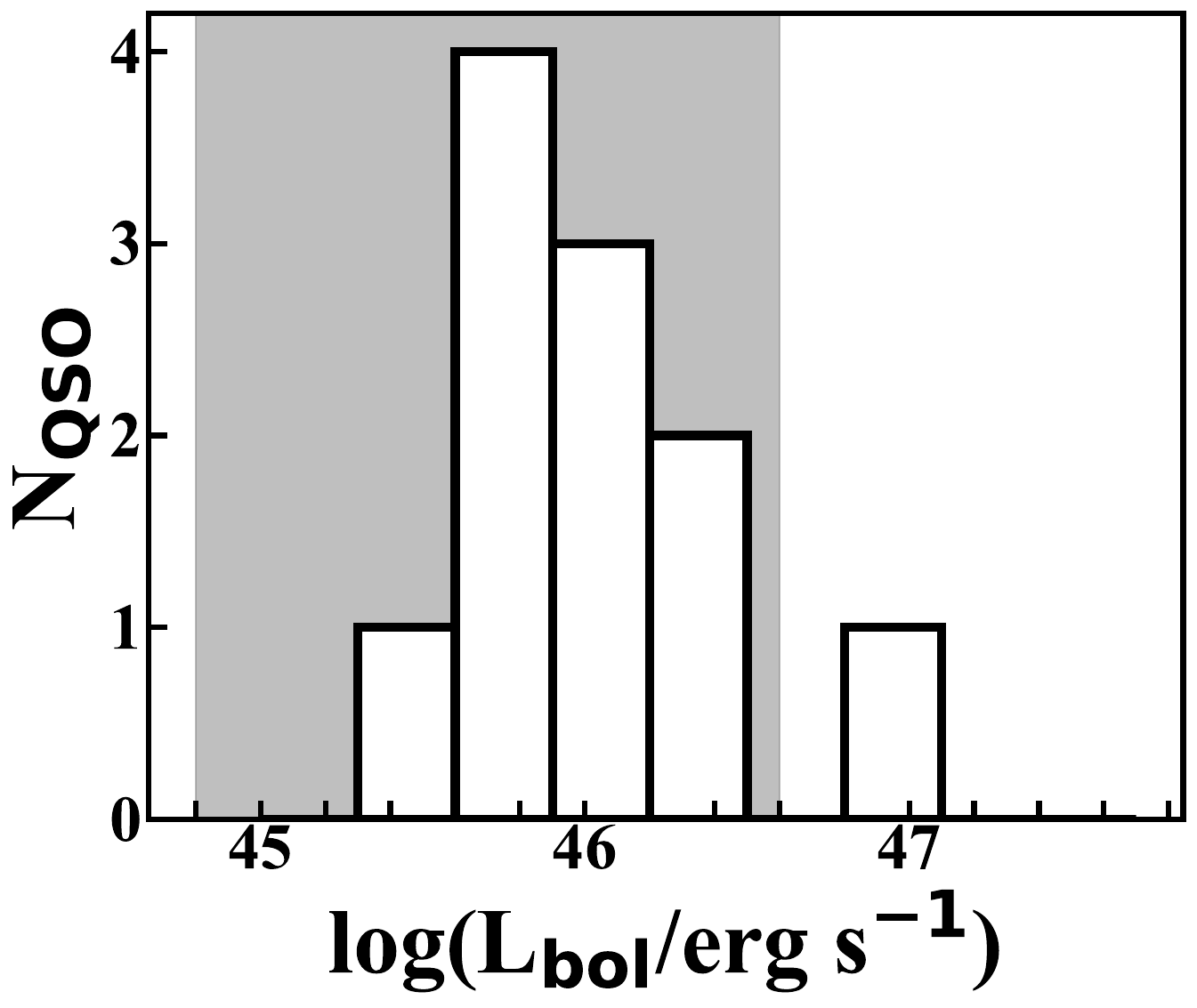}
    \caption{Bolometric luminosity ($L_{\rm bol}$) of proximate quasars in the \textit{Cosmic Himalayas}.
    All quasars exhibit $L_{\rm bol} \gtrsim 10^{45.5}~{\rm erg}~{\rm s}^{-1}$.
    The gray shaded area suggests the central 68\% scatter of $L_{\rm bol}$ in SDSS DR16 quasars spectroscopically determined by \citet{Wu+2022}.  
    }
    \label{fig:qso_Lbol}
\end{figure}

The field BOSS J0210+0052, also called J0210, is selected from the $10,000$ deg$^2$ SDSS/eBOSS database. 
It is traced by a rare group of nearby lines of sight toward background quasars that all exhibit strong Ly$\alpha$ absorption, as well as a significant quasar overdensity \citepalias{Liang+2021}.
The quasars are identified as type-1 AGNs with strong broad emission lines, based on spectra from the Baryon Oscillation Spectroscopic Survey (BOSS) of SDSS-III \citep{Dawson+2013} and the extended-BOSS (eBOSS) of SDSS-IV \citep{Dawson+2016}.
Moreover, spectra of background quasars from SDSS/eBOSS were utilized in \citetalias{Liang+2021} for selecting candidate fields for Subaru observations and analyzing Ly$\alpha$ absorption in IGM {\sc Hi} for galaxy–-IGM correlation studies.
The SDSS/eBOSS survey encompasses over 200,000 spectra of quasars down to $r=22$ across a survey area exceeding 10,000 deg$^2$, translating to a survey volume greater than 1 Gpc$^3$ at $z\sim2$ \citep{Dawson+2016}.
In focusing on the field J0210, this study examines both the proximate quasars within the redshift range $2.15\leq z \leq2.20$ and the background quasars at $z>2.20$ to construct 3D IGM H{\sc i} tomography maps, as detailed in Section \ref{sec:hi_tomography}.

\subsection{Subaru/Hyper Suprime-Cam Images}

The J0210 field was observed in \citetalias{Liang+2021} as a part of the MAMMOTH-Subaru project.
Observations were conducted to identify LAEs using HSC, 
a high-performance camera with a wide field of view (FoV; $\approx 1.5$ degrees in diameter), mounted at the prime focus of the 8.2-meter Subaru telescope atop Mauna Kea, Hawaii \citep{Miyazaki+2012, Miyazaki+2018}.
Broadband $g$-band and narrowband (NB) NB387 images ($\lambda_0$ = $3,862$ \AA, FWHM = 56 \AA) were taken by Subaru/HSC during queue observations between January 11 and January 20, 2018.
The total on-source exposure times were 165 minutes for NB387 and 40 minutes for the $g$-band, divided into eleven 900-second and four 600-second exposures, respectively. 
Frames observed under poor conditions, such as when the seeing exceeded FWHM $> 1.\arcsec3$ or tracking failed due to low transparency, were excluded.

Data reduction is detailed in \citetalias{Liang+2021} with Subaru/HSC pipeline \texttt{hscPipe} versions 5.4 and 6.6 \citep{Bosch+2017, Aihara+2019}.
As a result, the images achieved $5\sigma$ limiting magnitudes of NB387 = 24.36 and $g$ = 26.34.
The FWHMs of the point spread function (PSF) are 1.$\arcsec2$ and 0.$\arcsec9$ for NB387 and $g$-band.
During source extraction with \texttt{SExtractor} \citep{Bertin+1996}, regions saturated by the brightest stars or exhibiting high root-mean-square (RMS) noise due to CCD mosaicking failures were masked, shown as the gray shaded regions in the following sky maps.
The stacked NB387 image enables the detection of \lya~emission at redshifts of $z=2.177 \pm 0.023$, with the $g$-band image assessing the continuum levels of objects.
In total, 465 LAEs were selected by color excess in NB387, indicative of a rest-frame equivalent width for \lya~emission EW$_0 \geq$ 20 \AA.
The limiting Ly$\alpha$ luminosity reaches $L_{\rm Ly\alpha} \approx 2 \times 10^{42}~{\rm erg}~{\rm s}^{-1}$ in this LAE sample \citepalias{Liang+2021}.

This field is covered in the HSC Subaru Strategic Program's wide-field layers \citep[SSP;][]{Aihara+2018, Aihara+2019, Aihara+2022}, offering access to multiple broadband optical data sets with limits of $r<26.1$, $i<25.9$, $z<25.1$, and $Y<24.4$. 
Alongside photometry from SDSS, we utilized the HSC SSP third data release \citep[DR3;][]{Aihara+2022} to analyze the UV slopes and bolometric luminosities of quasars, as detailed in Section \ref{sec:sdss_qso}.

\section{Identifying The \textit{Cosmic Himalayas}} \label{sec:qso_lae}

\subsection{Quasar Overdensity}
\label{sec:sdss_qso}

\begin{figure*}[hbt!]
    \centering
    \includegraphics[width=0.8\linewidth]{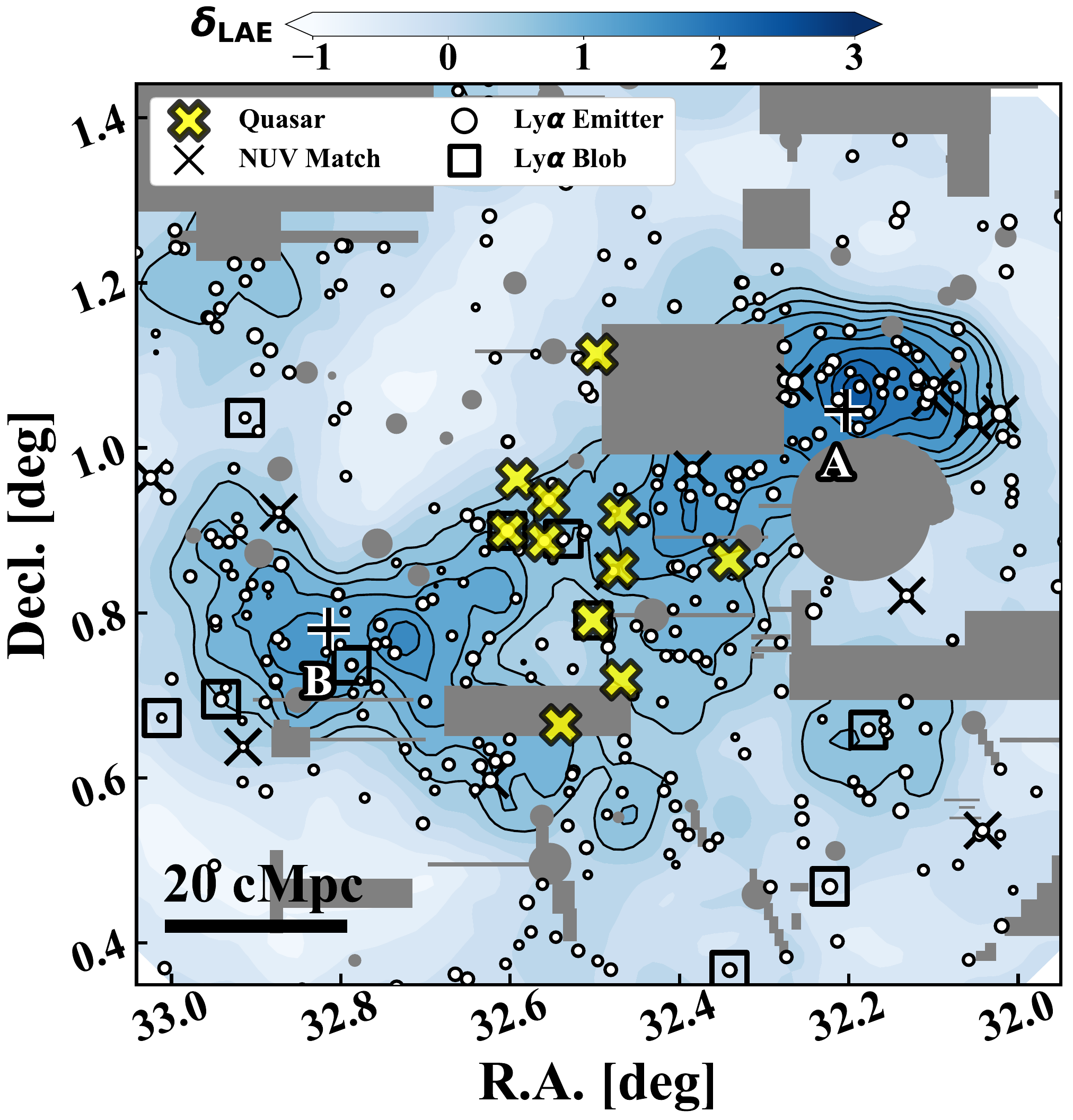}
    \caption{Overdensity map of LAEs, with black circles marking LAE positions, sized according to UV luminosity based on $g$-band. 
    The background contour represents LAE overdensity, $\delta_{\rm LAE}$, with bluer areas indicating higher densities. 
    Yellow crosses show SDSS/eBOSS quasars at $z=2.15-2.20$.
    Black squares and crosses highlight identified \lya~blobs (LABs) and the LAEs with NUV counterparts from GALEX MIS survey (NUV-LAEs).
    The black pluses mark the median centroids of member LAEs in Node-A (NW) and Node-B (SE), respectively, which are the two overdensities traced by LAEs. 
    The gray shaded regions indicate the masks due to saturation by bright stars or high RMS noise with CCD mosaicking failures.} 
    \label{fig:lae_od}
\end{figure*}

We report on a unique cosmological structure in the J0210 field, identified through SDSS quasars and dubbed the \textit{Cosmic Himalayas}. 
Eleven proximate quasars are situated in 
a volume of approximately $(40 {\rm cMpc})^3$ within J0210.
We visually inspect all their spectra and confirm that they are indeed type-1 QSO with broad emission.
These quasars represent a group of luminous AGNs, with M$_{\rm UV}\lesssim-23$ at $z=2.167-2.200$, derived from their SDSS $u$-band magnitudes. 
Their redshifts are spectroscopically estimated by the SDSS pipeline based on PCA fitting, and we have also visually confirmed their reliability. 
Their spectroscopic redshifts and $i$-band magnitudes, a primary reference for SDSS/eBOSS target selection \citep{Dawson+2016}, are listed in Table \ref{tab:qso_info}.

The quasar concentration in the \textit{Cosmic Himalayas} represents the highest density peak of SDSS quasars across the SDSS/eBOSS survey area of 10,000 deg$^2$ sky in the redshift range $z=2.15-2.20$, as illustrated in Figure \ref{fig:qso_sky} (a).
The distribution of SDSS quasar number counts is depicted in Figure \ref{fig:qso_sky} (b), with an average quasar number of 0.36 per $(40 {\rm cMpc})^3$ volume across the survey area and a standard deviation of 0.63 from a Gaussian fit to the distribution.
Thus, the quasar overdensity, 
which is defined as the number density excess compared to the mean over the SDSS survey area $\delta_{\rm QSO} = N_{\rm QSO} / \left<N_{\rm QSO}\right> - 1$, is calculated to be $\delta_{\rm QSO}\approx30$. 
This $\delta_{\rm QSO}$ reaches a significance of $16.9 \sigma$, marking it as the most significant AGN overdensity discovered to date at $z>2$ on a similar or larger scales. 
This concentration exceeds by a factor of two to three the previously identified extreme cases, such as BOSS1244 or BOSS1441, which contained 4 or 5 quasars at a similar rest-UV detection limit of $M_{\rm UV}=-23$ within a comparable volume \citep{Cai+2017b, Shi+2021}.

For reference, we estimate the bolometric luminosities $L_{\rm bol}$ of the eleven quasars.
Utilizing the SDSS $ugriz$ and HSC SSP $grizY$ bands, their UV continuum slopes $\beta_{\rm QSO}$ are estimated by fitting a power-law continuum.
The slope $\beta_{\rm QSO}$ aids in predicting the monochromatic continuum luminosity at rest-frame 1350 \AA, denoted as $L_{1350}$. 
With the correction factor from \citet{Shen+2011}, we calculate the bolometric luminosity as follows:
\begin{equation}
    L_{\rm bol} = 3.81 \times L_{1350}.
\end{equation}
Given that this conversion introduces a 50\% uncertainty, which dominates other sources of error, we adopt 50\% of the estimated values as the associated uncertainties.
These results, detailed in Table \ref{tab:qso_info} and depicted in Figure \ref{fig:qso_Lbol}, show that all eleven quasars 
exhibit $L_{\rm bol}\gtrsim10^{45.5}$ erg s$^{-1}$, indicating their status as part of a relatively luminous population among DR16 quasars \citep{Wu+2022}.
We label the eleven SDSS quasars with IDs ranked by their $L_{\rm bol}$, and the quasars from CH-Q01 to CH-Q11 are those from the most to the least.
We note that the measurements based on SDSS and HSC photometry are consistent with previous literature using SDSS spectra within 1\% for CH-Q01 and CH-Q02; within 0.3 dex for all quasars, including the faintest ones \citep{Wu+2022}. 

\subsection{LAE Overdensities and Properties}

In \citetalias{Liang+2021}, Subaru/HSC NB387 observations were performed on J0210 to select the coeval LAEs -- a young galaxy population with strong Ly$\alpha$ emission, at redshift $z=2.177\pm0.023$.
While located at the same redshifts as the quasars, these LAEs also well overlap the SDSS quasars in the \textit{Cosmic Himalayas} in the projected sky, as shown in Figure \ref{fig:lae_od}. 
Notably, the LAE candidates construct a cosmological filament with an end-to-end size of over $100$ cMpc. 
This cosmological filament intersects the quasar overdensity and links two cosmic nodes, i.e., the overdensity of LAEs in the projected sky.
The northwestern (NW) node is labeled as Node-A, and the southeastern (SE) node is labeled as Node-B.
The centroids of the two nodes, 
iteratively derived from the median coordinates of member galaxies as defined below, 
are labeled as points A and B, respectively.

The LAE overdensities at Node-B and Node-A exhibit significances of approximately $4\sigma$ and $6\sigma$, respectively, with peak overdensities of $\delta_{\rm LAE}\approx1.5$ and $\approx2.5$ on a 20 cMpc scale.
Assuming each node has a spherical shape with similar dimensions in both the line-of-sight (LOS) and transverse directions, and that they will eventually collapse into a single virialized structure by $z=0$, their predicted halo masses at $z=0$ are M$_h(z=0) \approx 10^{14.3}~{\rm M}{\odot}$ for Node-A and M$_h(z=0) \approx 10^{14.8}~{\rm M}{\odot}$ for Node-B. These estimates are based on simulations of galaxy overdensities on a comparable 25 cMpc scale \citep{Chiang+2013}.
This suggests both nodes are likely progenitors of Virgo-type galaxy clusters.
Contrary to expectations, the quasar overdensity is positioned midway along the filament, not near either LAE node but with an offset of about 25 cMpc.

Within J0210, we further examine the properties of LAEs at Node-A and Node-B, containing 78 and 76 member LAEs, respectively. 
These members are defined by their separation from points A or B, being less than $15$ $h^{-1}$cMpc, or approximately $22$ cMpc.
We set the criterion as $15$ $h^{-1}$cMpc because this is a typical size to enclose the entire outskirts of protoclusters for meaningful comparison to simulations \citep[e.g.,][]{Chiang+2013}, a potential scale to have the most significant galaxy--IGM correlation signal \citep[e.g.,][and \citetalias{Liang+2021}]{Cai+2016}, and a comparable scale to the spatial resolution of IGM tomography map constructed in Section \ref{sec:hi_tomography}. 

Initially, we employ the observed $g$-band magnitudes as proxies for the UV luminosity of LAEs to conduct comparisons between the two nodes. 
In Figure \ref{fig:lae_od}, the sizes of circles are encoded by LAEs' $g$-band magnitudes, with larger circles indicating higher UV luminosities. 
Visually, LAEs in Node-A appear systematically more luminous than those in Node-B. 
This observation is statistically supported by a two-sample Kolmogorov–Smirnov (K-S) test, yielding a $p$-value of approximately $10^{-4}$, which quantitatively confirms the significant difference in UV luminosity distributions between the two nodes.

Secondly, we cross-matched LAE candidates with near-UV (NUV; effective wavelength $\lambda_{\rm eff}=2,314$ \AA) detections down to $m_{\rm AB}\approx22.7$ from the GALEX Release (GR) 6/7 of the Medium-depth Imaging Survey \citep[MIS;][]{Bianchi+2017}. 
The matching aperture is set to 2\arcsec, a size smaller than the PSF of GALEX NUV images. 
This selection enables the exclusion of false matches with high-resolution HSC images upon visual inspection.
Only sources with NUV $<23$ and errors $<0.5$ mag are accepted. 
Given the Lyman break at $z>2.15$ falls redward of the NUV wavelength range, no detection is typically expected for the star-forming galaxies within the shallow GALEX NUV images for our continuum-faint LAEs, as per the LBG selections \citep{Ly+2009, Haberzettl+2012}. 
Out of 465 LAEs, 13 were identified with NUV detection, referred to as NUV-LAEs, and are indicated by thin black crosses in Figure \ref{fig:lae_od}. 
Of these, 6 belong to Node-A and 2 to Node-B, corresponding to fractions of 8\% and 3\% member LAEs, respectively, with NUV detection in each node. 
These NUV-LAEs could represent either low-$z$ interlopers or faint AGN candidates, with a notable clustering in Node-A and a tendency to avoid Node-B.

A systematic search for Ly$\alpha$ blobs (LABs), galaxies exhibiting luminous and extended Ly$\alpha$ emission, was conducted using data from the MAMMOTH-Subaru survey, including the J0210 field \citep[][M. Li et al. in prep.]{ZhangHB+2023}. 
The LABs are selected when their \lya~isophotal area $A_{\rm iso}$ is $3\sigma$ above the sequence of point-like sources at the given \lya~luminosity $L_{\rm Ly\alpha}$ and $A_{\rm iso}$ is at least over about 15 arcsec$^2$. 
From the LAB catalog, we find ten LABs in J0210, and they are indicated as the black squares in Figure \ref{fig:lae_od}.
Notably, six LABs distribute eastward of the quasar overdensity towards Node-B, two of which (CH-Q05 and CH-Q08) are also identified as LABs, while the west lobe of the LAE filament towards Node-A lacks any proximate LABs.  
LABs are thought to inhabit gas-rich environments, attributed to Ly$\alpha$ resonant scattering, cooling radiation, 
or recombination radiation 
as potential mechanisms producing extended \lya~emission \citep{Tumlinson+2017, Ouchi+2020}.
This suggests that Node-B may be at an earlier stage of stellar mass buildup compared to Node-A.

These characteristics indicate that the two cosmic nodes of LAEs may represent independent substructures with distinct evolutionary histories or at different evolution stages.  
Specifically, Node-A appears to be a more mature large-scale structure compared to Node-B. 
Intriguingly, the overdensity of proximate quasars is situated at the transition point of LAE properties.

\section{3D IGM {\sc Hi} Tomography Map} 
\label{sec:hi_tomography}

\subsection{Methodology}
\label{sec:hi_tomo_method}

\begin{figure*}[hbt!]
    \centering
    \includegraphics[width=\linewidth]{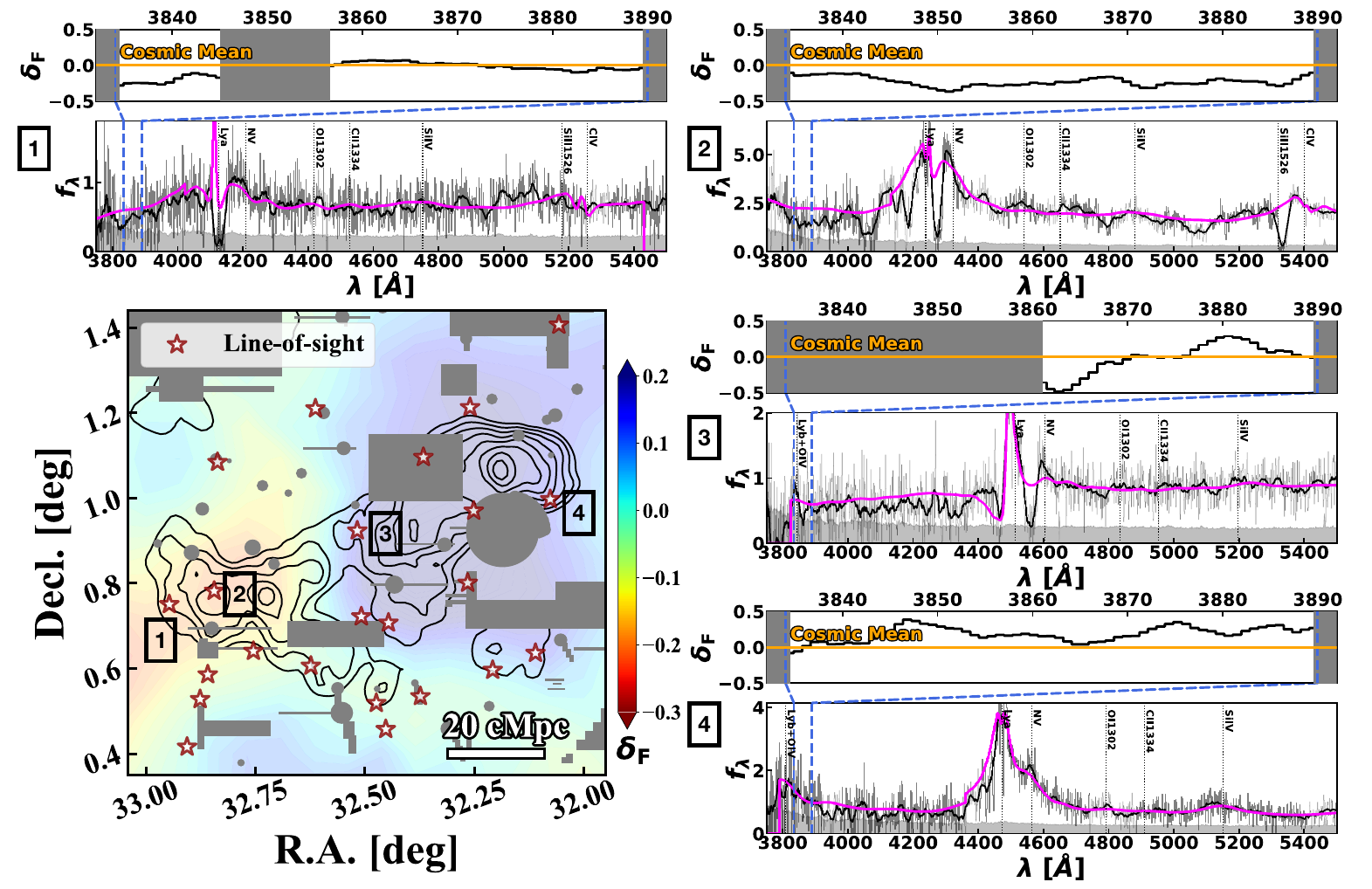}
    \caption{Examples of the MF-PCA fitting of quasar spectra to derive Ly$\alpha$ transmission fluctuations for reconstructing 3D {\sc Hi} tomography maps.
    The lower left panel is a collapsed map of {\sc Hi} absorption within redshift $2.15<z<2.20$.
    Red stars mark the LoSs used for tomography, with labeled ones corresponding to the examples in other panels.
    Sequentially from the upper left to the lower right panels, MF-PCA fits of quasar spectra along the LAE filament are illustrated.
    Each example highlights the NB387 redshift/wavelength coverage with blue dashed vertical lines. 
    In each spectrum's lower panel, thin and thick black lines represent the original and $15h^{-1}{\rm cMpc}$-smoothed spectra, respectively, and the magenta line denotes the MF-PCA fit. 
    The upper panel of each spectrum shows the $15h^{-1}{\rm cMpc}$-smoothed Ly$\alpha$ transmission fluctuation $\delta_F$ within the focused wavelength range, with the orange line indicating the cosmic mean Ly$\alpha$ transmission at the relevant redshift.
    Grey-shaded regions are masked out for analysis, with criteria detailed in text.}
    \label{fig:hi_trans_method}
\end{figure*}

The overdensities of neutral hydrogen in IGM can be revealed by 3D IGM tomography, which is reconstructed by the Ly$\alpha$ forest imprinted in the continuum of background sources.
We utilize the SDSS quasar spectra to reconstruct a coarse IGM tomography map with a resolution of 15 $h^{-1}$cMpc.
The reconstruction procedures of IGM tomography follow \cite{Sun+2023}, which is further based on the methodology proposed in the COSMOS Lyman-Alpha Mapping And Tomography Observations \citep[CLAMATO;][]{Lee+2016, Lee+2018, Horowitz+2022}. 
We briefly review the essential workflow in the following paragraphs.

The IGM map is made from {\sc Hi} absorption left on the spectra of multiple background quasars in the sky. 
First, damped Ly$\alpha$ absorption systems (DLAs), characterized by high-column density hydrogen gas with $N_{\rm HI}>2\times10^{20}~{\rm cm}^{-2}$ likely associated with galaxies, are masked in the spectra to avoid biasing the estimate of IGM H{\sc i} Ly$\alpha$ absorption. 
The masked wavelength range is defined by the equivalent width (EW) of the DLAs and requires a continuum-to-noise ratio (C/N) greater than 1.0. 
Additionally, potential metal absorption lines, such as Si{\sc iv} $\lambda 1062$, N{\sc ii} $\lambda 1084$, N{\sc i} $\lambda 1134$, and C{\sc iii} $\lambda 1176$, are masked in the intrinsic quasar spectrum using a $10$ \AA~mask in the observer's frame \citep{Lee+2012}.

\begin{figure*}[t]
    \centering
    \includegraphics[width=0.8\linewidth]{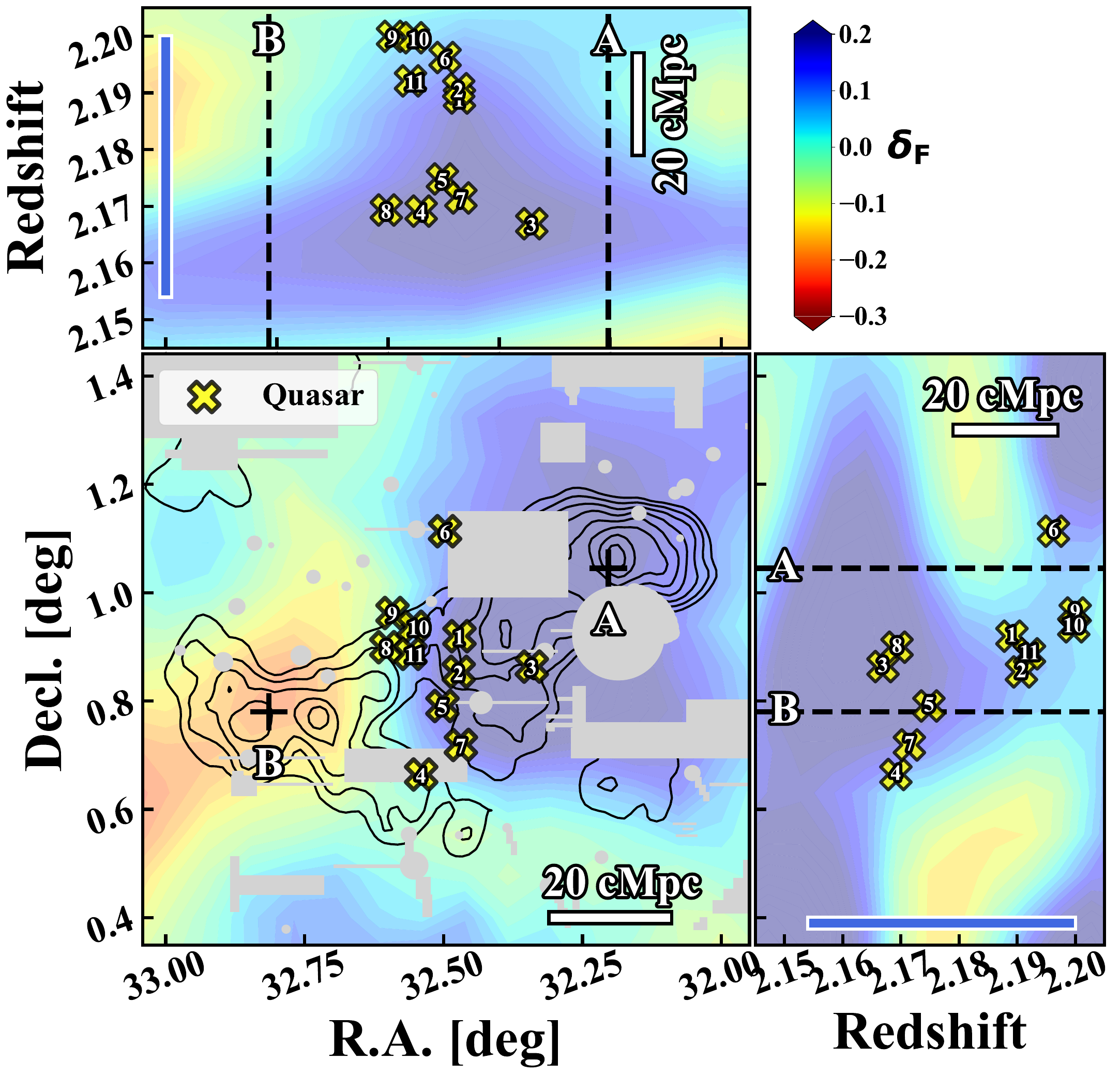}
    \caption{Collapsed H{\sc i} absorption maps from three perspectives over about $60$ cMpc:
    (a) {\it Lower left:} The R.A.--Decl. plane, collapsed across $2.15<z<2.20$;
    (b) {\it Upper left:} The R.A.--$z$ plane, collapsed within $0.6<{\rm Decl.}<1.2$;
    (c) {\it Lower right:} The $z$--Decl. plane, collapsed for $32.2<{\rm R.A.}<32.8$.
    The colored contour in the background represents Ly$\alpha$ transmission fluctuation $\delta_F$ after a median stacking over the specific dimensions, with a bluer color indicating higher $\delta_F$, indicative of lower {\sc Hi} density.
    Yellow crosses are the proximate quasars, the same as Figure \ref{fig:lae_od}, with numbers indicating their IDs.
    Blue bars represent the redshift range of LAEs identified using NB387.
    }
    \label{fig:hi_trans_map}
\end{figure*}

Second, mean-flux regulated principal component analysis \citep[MF-PCA;][]{Lee+2016, Suzuki+2005} is utilized to model the intrinsic continuum in the Ly$\alpha$ forest wavelength range (i.e., rest-frame 1,040–1,185 \AA) from the spectrum redward of the Ly$\alpha$ emission.
Following a conventional least-squares PCA fit to the spectra of background quasars, the method includes an adjustment step, aligning the cosmic mean Ly$\alpha$ transmission to correct the flux uncertainty of SDSS/eBOSS spectra blueward of the Ly$\alpha$ emission \citep{Lee+2012, Becker+2013}. 
The cosmic mean Ly$\alpha$ transmission follows the empirical relation, $\langle F(z)\rangle_{\rm cos}=\exp \left[-0.001845(1+z)^{3.924}\right]$, approximating $0.84$ at $z\approx2.2$ \citep{FG+2008}. 
Spectra (1) -- (4) in Figure \ref{fig:hi_trans_method} exemplify the fitted spectra and continua for four LOSs along the LAE filament.

To reconstruct the H{\sc i} tomography maps using these LoSs, we define a grid-based Ly$\alpha$ forest fluctuation as:
\begin{equation}
 \delta_{F} = F_{\rm Ly\alpha}/\langle F(z) \rangle_{\rm cos} - 1.
\end{equation}
Here, $F_{\rm Ly\alpha}$ represents the Ly$\alpha$ forest transmission at each pixel on the background quasar spectrum, calculated as $F_{\rm Ly\alpha}=f_{\rm obs}/f_{\rm int}$, where $f_{\rm obs}$ and $f_{\rm int}$ are the observed spectrum and the intrinsic continuum from MF-PCA, respectively.
Note that $\delta_{F,~{\rm lim}}\approx0.2$ is the physical upper limit for fully ionized IGM with $F_{\rm Ly\alpha}=1$, and the measured values higher than this limit will be replaced by $\delta_{F,~{\rm lim}}$. 
A Wiener filter is then applied to the LoSs to reconstruct the grid-based H{\sc i} tomography map across the entire 3D comoving volume covering the J0210 field. This volume spans $30.5<{\rm R.A.}<34.5$ and $-1.1<{\rm Decl.}<2.9$ in the redshift range $2.10<z<2.25$, consisting of $52\times52\times27$ Cartesian grids with a cell size of $5.114~h^{-1}{\rm cMpc}$ along each axis. The H{\sc i} tomography map is created over a broader volume than the J0210 region of interest to mitigate boundary effects.

Initially, 455 spectra of background quasars  
from SDSS/eBOSS DR18 were available for reconstructing the H{\sc i} tomography map. 
Of these, 428 were chosen as LoSs after visual inspection post MF-PCA fitting to remove those of poor quality. 
Ultimately, 242 LoSs, with suitable C/N ratios and redshift $z>2.23$, were utilized to reconstruct the entire volume. 
Among them, 23 LoSs fall within the J0210 HSC FoV providing critical information for our study, and are depicted as red stars in Figure \ref{fig:hi_trans_method}. 
A summary of the reconstructed H{\sc i} tomography, a collapsed map by stacking the $\delta_{\rm F}$ over redshifts $z=2.15-2.20$, is presented as the background contour color-coded in Figures \ref{fig:hi_trans_method} and \ref{fig:hi_trans_map}.
The $\delta_F$ values at QSO positions are also presented in Table \ref{tab:qso_info}.
Due to the coarse sampling of the LoSs, the uncertainty in $\delta_F$ is primarily driven by cosmic variance, which can only be mitigated through statistical analysis.
In the following sections, we adopt a 50\% variation as the uncertainty to illustrate its potential impact on the statistical results when necessary.

\subsection{IGM Tomography Results}
\label{sec:tomo_result}
\subsubsection{Average IGM Properties Over $2.15 \leq z \leq 2.20$}
\label{sec:tomo_result_average}

\begin{figure*}[hbt!]
    \centering
    \includegraphics[width=\linewidth]{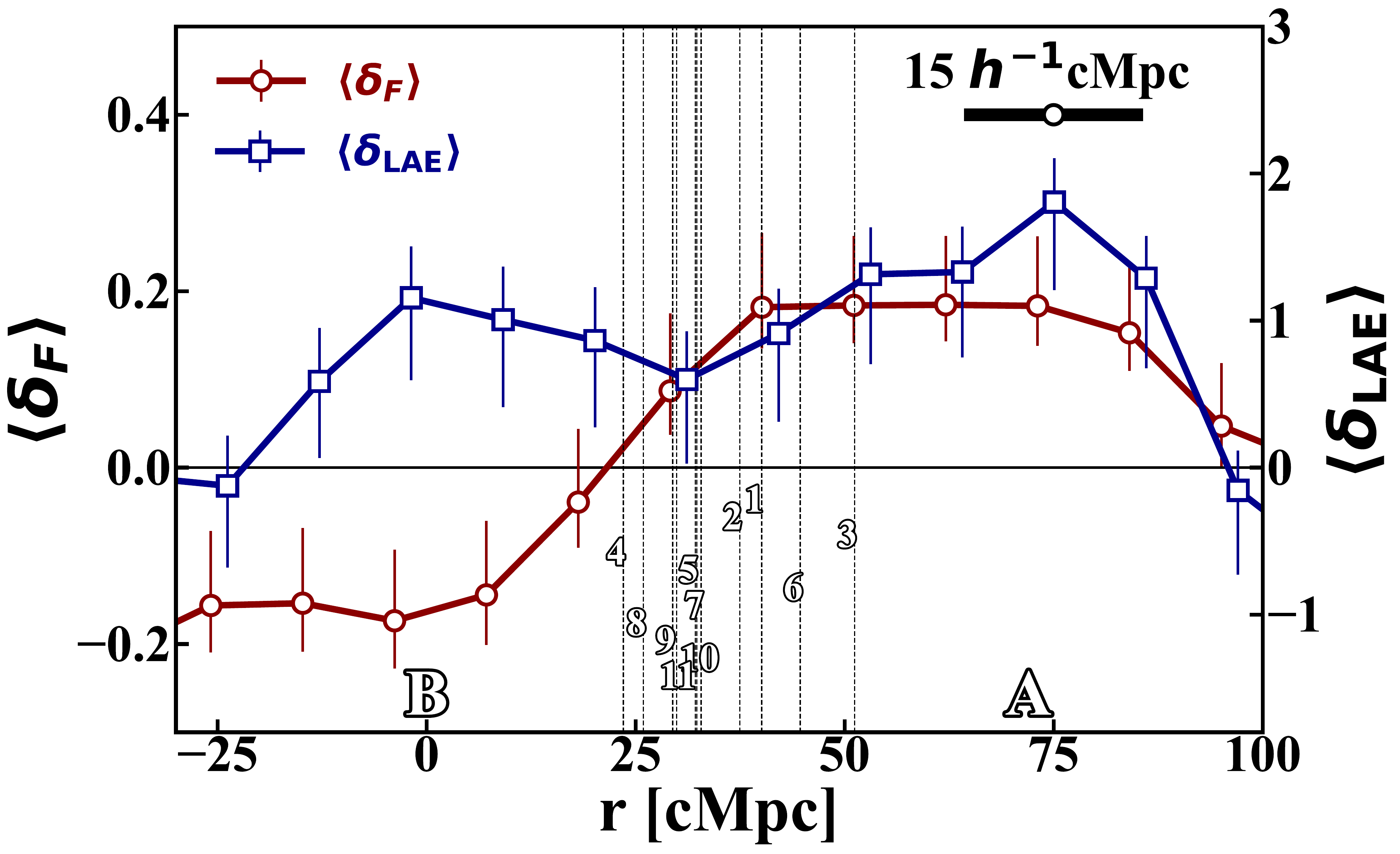}
    \caption{Profiles of averaged H{\sc i} absorption and LAE overdensity along the large-scale filament. Dotted vertical lines, labeled with numbers, mark the positions of proximate quasars by their IDs. Red circles indicate averaged H{\sc i} absorption, $\langle\delta_F\rangle$, and blue squares denote averaged LAE overdensity, $\langle\delta_{\rm LAE}\rangle$. Error bars reflect the 16\%--84\% scatter from 1,000 Monte Carlo simulations involving random placements. The origin for the X-axis is designated as point B.}
    \label{fig:hi_trans_profile}
\end{figure*}

The IGM tomography map is constructed in 3D.
For a quantitative analysis, we begin by examining collapsed maps on the R.A.--Decl. plane, where $\delta_{\rm F}$ is averaged over the redshift range $2.15 \leq z \leq 2.20$ using median stacking, as depicted in the main panel of Figure \ref{fig:hi_trans_map}. 
The top and side views, shown in the upper and right-hand panels, illustrate the quasar distribution along the redshift space. 
It's important to note that these collapsed maps average over dimensions of approximately $60$ cMpc each, determined by the redshift coverage of NB387--Ly$\alpha$. 
Therefore, the collapsed $\delta_{\rm F}$ values do not imply a direct connection to individual quasars but rather indicate a statistical correlation on a scale of about 60 cMpc.

The IGM tomography map indicates that the hydrogen gas in Node-B is more neutral than the cosmic average, with $\delta_{\rm F}\approx-0.1$, whereas Node-A is situated in an almost fully ionized region, with $\delta_{\rm F}\approx0.2$.  
Notably, a bimodal characteristic in the ionization state of the IGM also manifests along the LAE filament, reinforcing the distinct galaxy properties observed between the nodes.

The quasar overdensity exhibits substructures in redshift space, with the proximate quasars divisible into two groups: one comprises (CH-)Q03, Q04, Q05, Q07, and Q08 at $z\approx2.17$, and the other includes (CH-)Q01, Q02, Q06, Q09, Q10, and Q11 at $z\approx 2.19-2.20$. Each group occupies a volume of $\left(20~{\rm cMpc}\right)^3$, with a separation of approximately $20$ cMpc between the two.
A visual inspection reveals that most proximate quasars align with the interface between ionized and neutral regions. 
The exception is CH-Q03, which is situated in a fully ionized region with $\delta_{\rm F}\approx0.2$ across all dimensions, maintaining a distance of at least $15$ cMpc from the $\delta_{\rm F}=0$ boundary.

For a more quantitative analysis, a Ly$\alpha$ transmission fluctuation profile is constructed along the LAE filament from Node-B to Node-A, using point B as the origin. 
The filament A--B is inclined at 23.46 degrees. 
The width of the calculating box, perpendicular to filament A--B, is set to 15 $h^{-1}$cMpc, aligning with the smoothing scale of the IGM tomography map. 
This profile is depicted by red circles with error bars and a curve in Figure \ref{fig:hi_trans_profile}, while the locations of proximate quasars are indicated by vertical dotted black lines with ID labels. Additionally, the LAE overdensity profile within the same volume is represented by blue squares with error bars and curves. 
The uncertainties for both profiles are derived from the 16\% and 84\% percentiles in Monte Carlo simulations. 
These simulations are conducted by randomly placing the calculation boxes 1,000 times with varying orientations across the entire IGM tomography map beyond the J0210 region, covering a volume with R.A. in [30.5, 34.5] deg, Decl. in [-1.1, 2.9] deg, and $z$ in [2.10, 2.15].
As a result, the uncertainty estimate inherently includes the effect of cosmic variance due to LSS.

This profile quantitatively highlights the transition observed in our visual inspections, depicting the ionization stage shift of the IGM hydrogen from ionized gas in Node-A to neutral H{\sc i} in Node-B along the LAE filament. 
The proximate quasars exhibit a $\sim 25$ cMpc offset from the peaks of either the Ly$\alpha$ absorption or Ly$\alpha$ transmission in the IGM. 
These quasars are situated near the ionization frontiers where $\langle\delta_{\rm F}\rangle\approx0$, predominantly within relatively ionized regions.

The signal indicating an ionization stage transition between Node-A and Node-B, marked by a decrement in $\Delta \delta_{\rm F} \approx 0.34 \pm 0.11$, is significant on the scale of $50~h^{-1}{\rm cMpc}$.
This observation exceeds the limitations imposed by the IGM tomography map's spatial resolution, which is reconstructed from background SDSS quasar spectra with a mean LoS sampling separation of $D_{\perp}\approx 15~h^{-1}{\rm cMpc}$. 
The observed change distinctly stands out from general fields, where $\Delta \delta_{\rm F} \leq 0.03$, utilizing similar SDSS LoS samples and methodology \citep{Sun+2023}.
Note that the Wiener filter inherently smooths the intrinsic signal, particularly in regions with low-S/N or low-density spectra. 
As a result, the actual increment may be even more significant than observed.

\subsubsection{Inspection on Individual Redshifts}
\label{sec:tomo_result_slice}

\begin{figure*}[hbt!]
    \centering
    \includegraphics[width=0.325\linewidth]{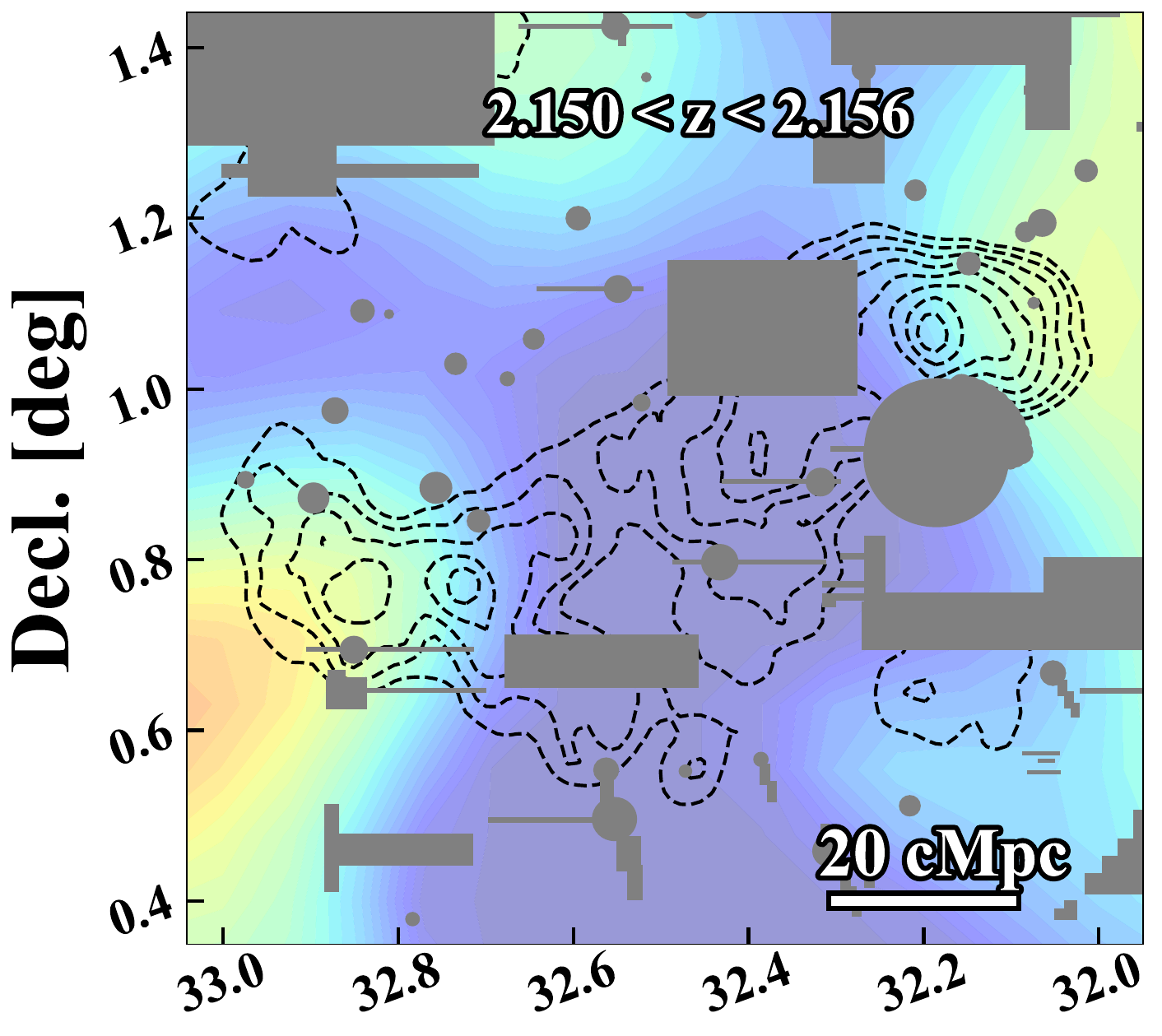}
    \includegraphics[width=0.3\linewidth]{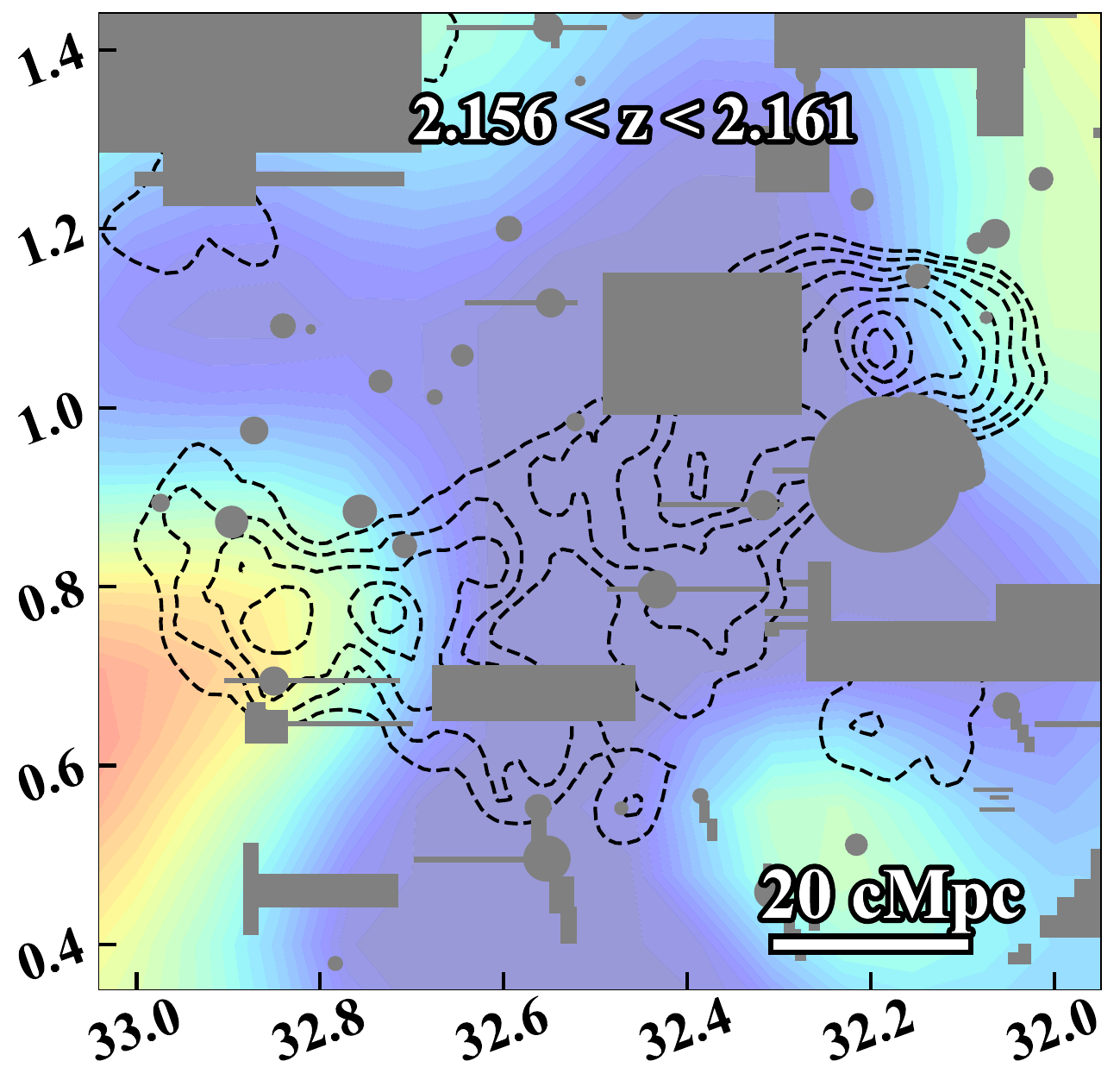}
    \includegraphics[width=0.343\linewidth]{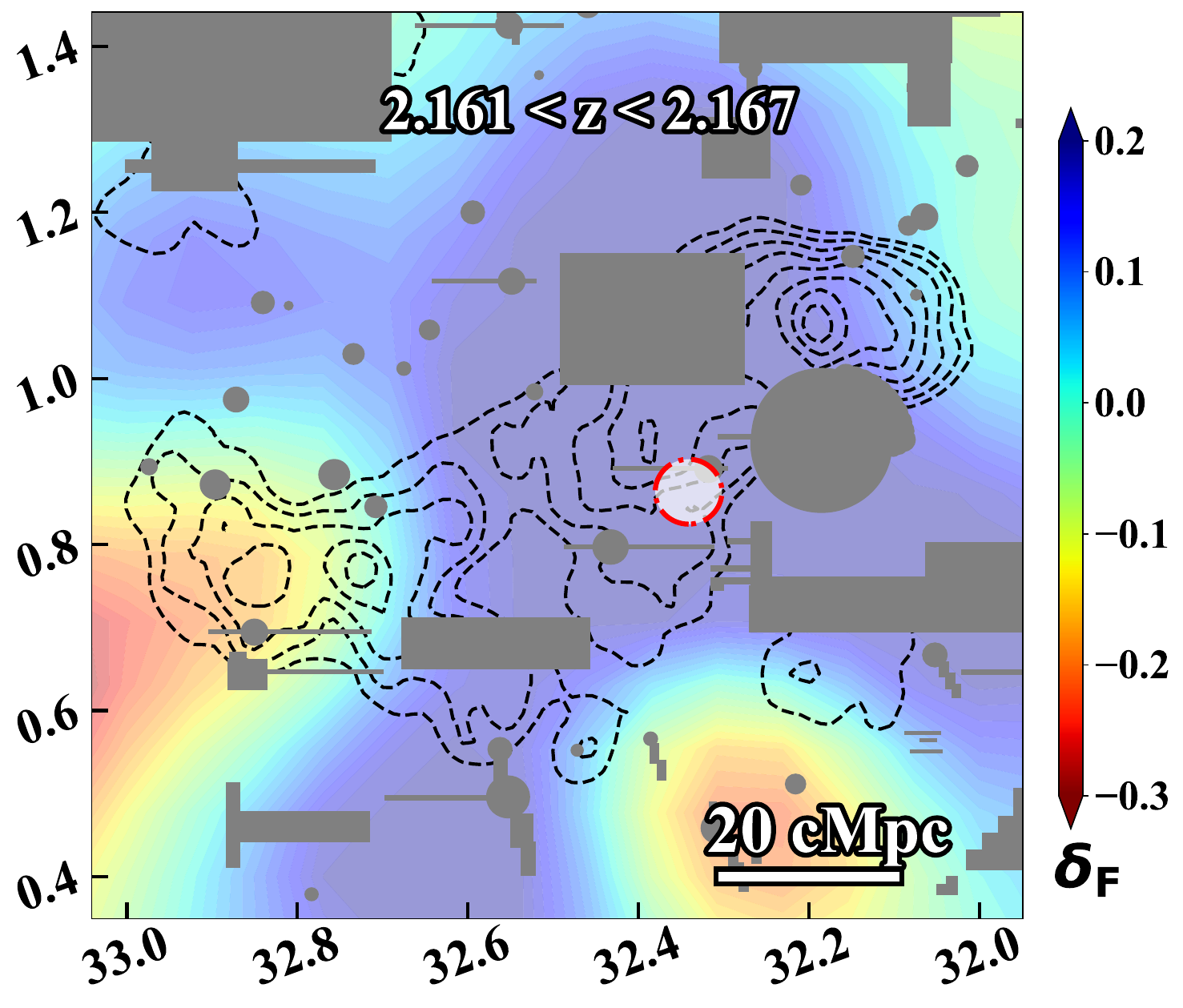}
    \includegraphics[width=0.325\linewidth]{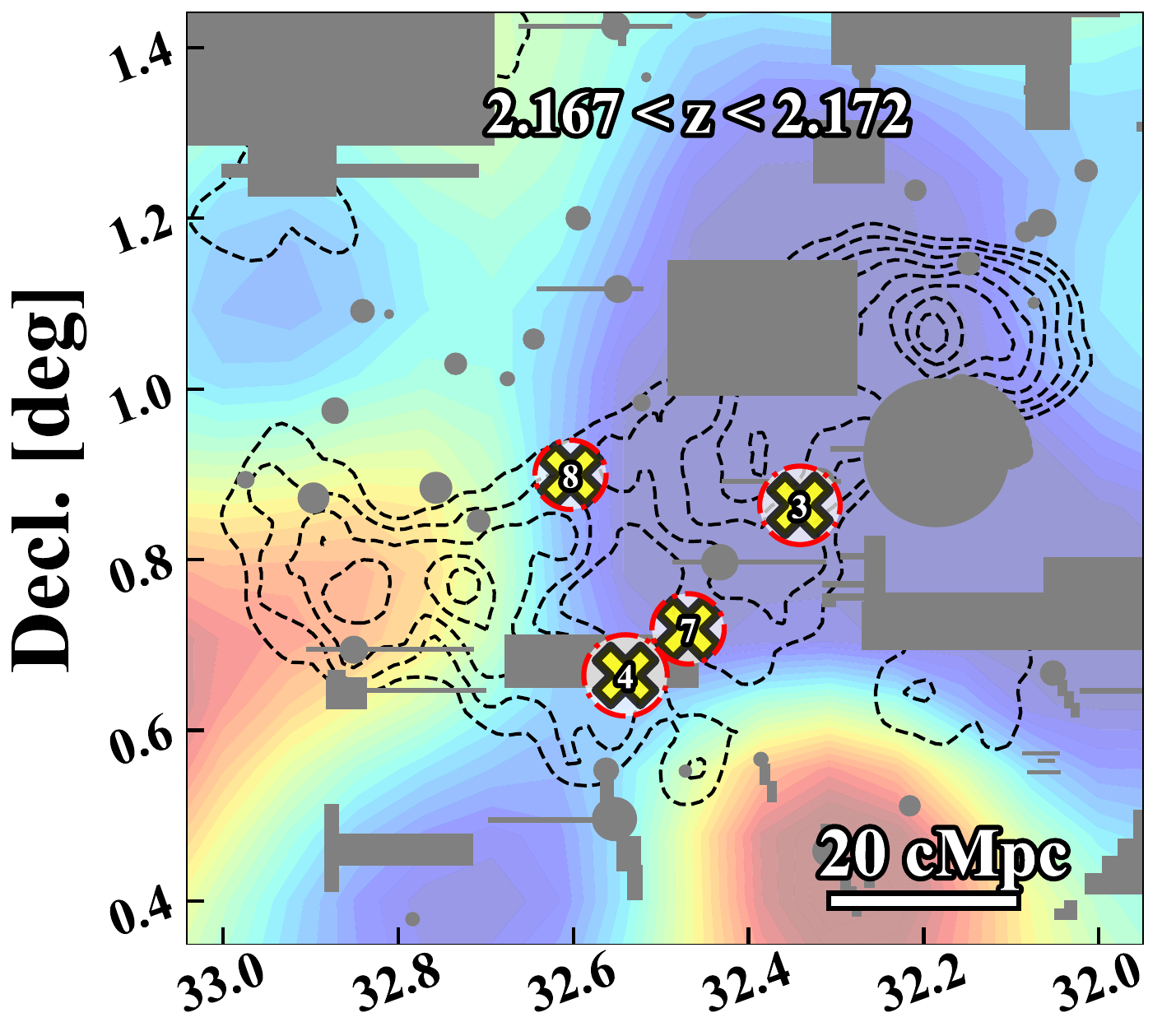}
    \includegraphics[width=0.3\linewidth]{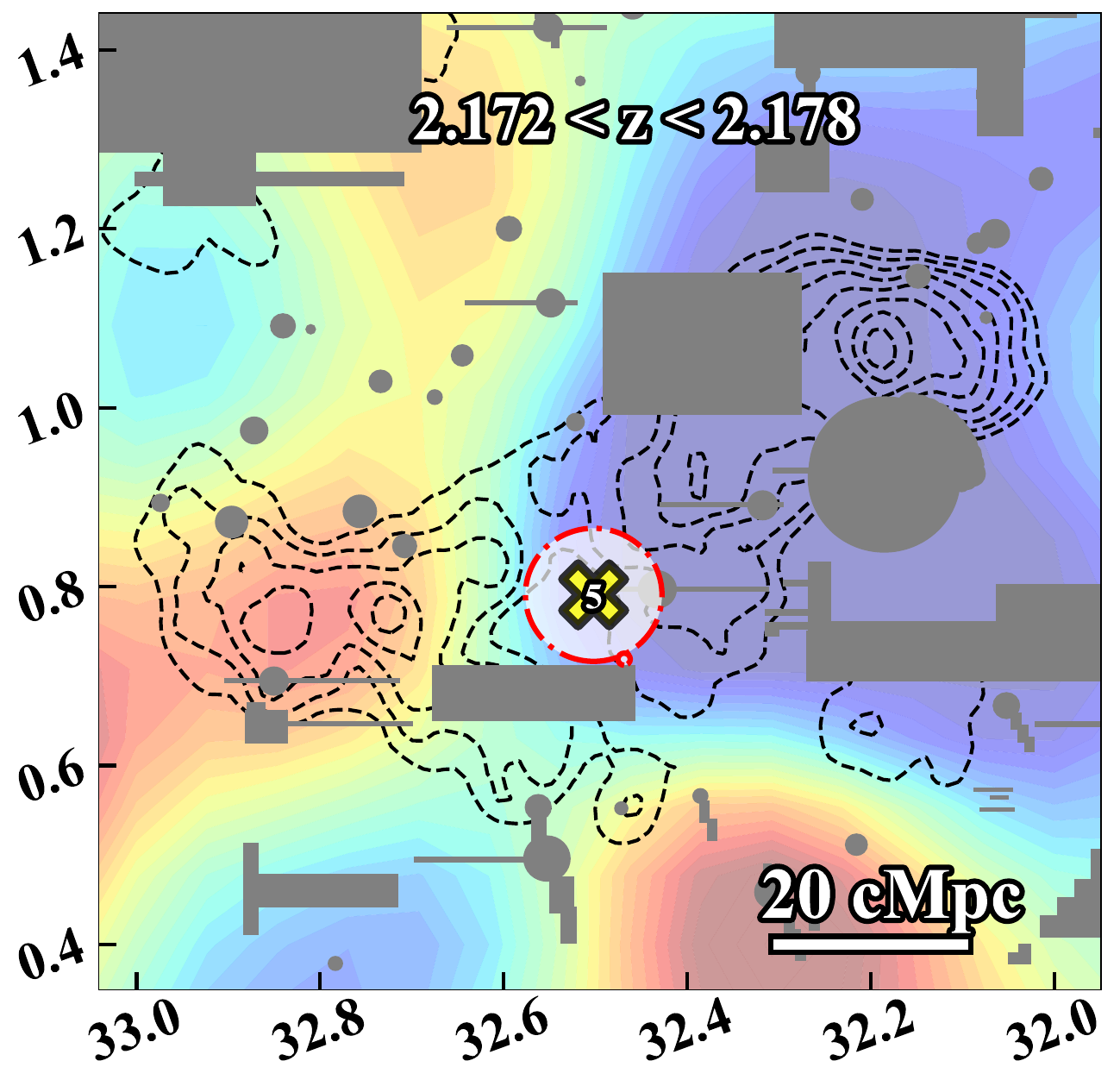}
    \includegraphics[width=0.343\linewidth]{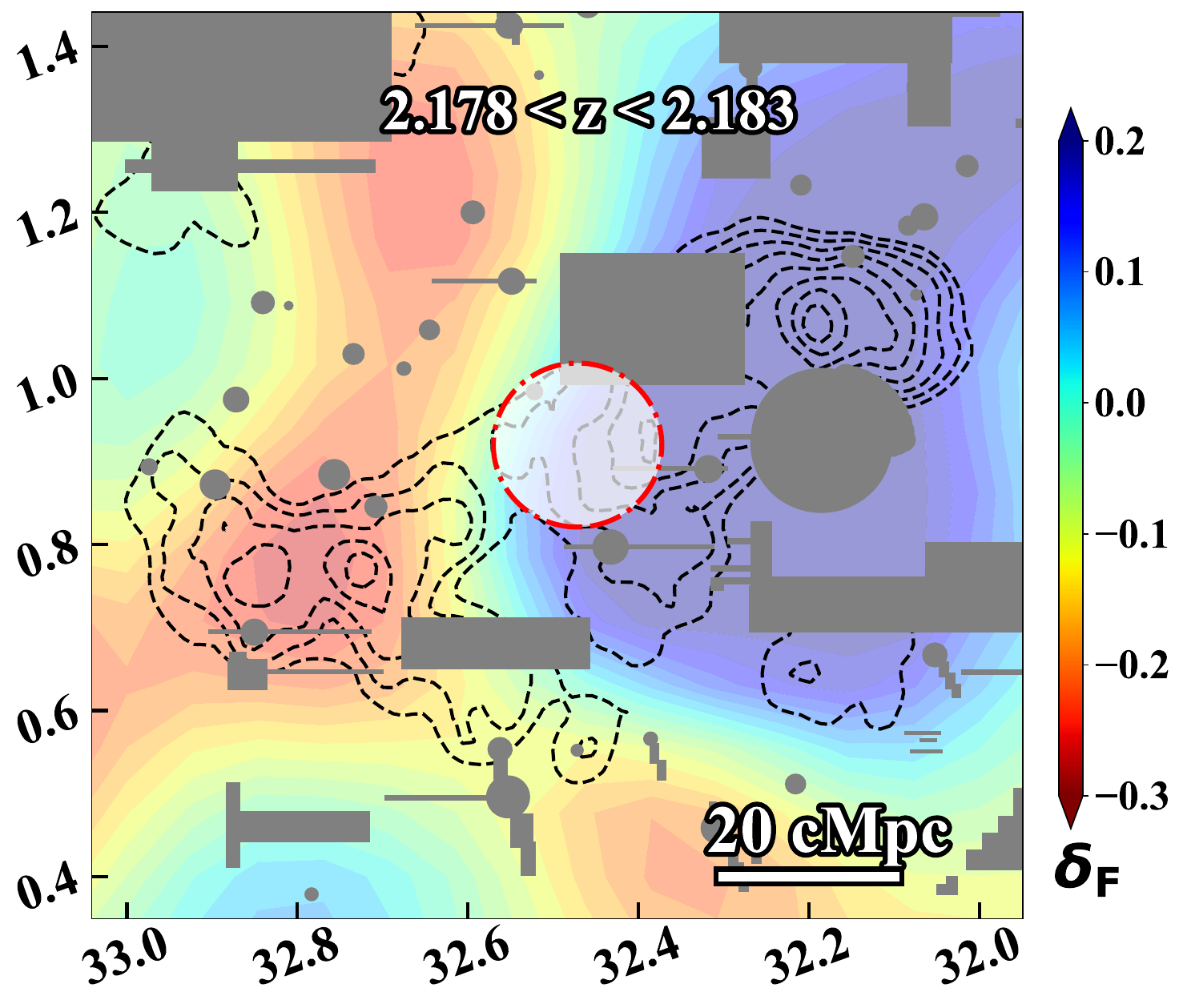}
    \includegraphics[width=0.325\linewidth]{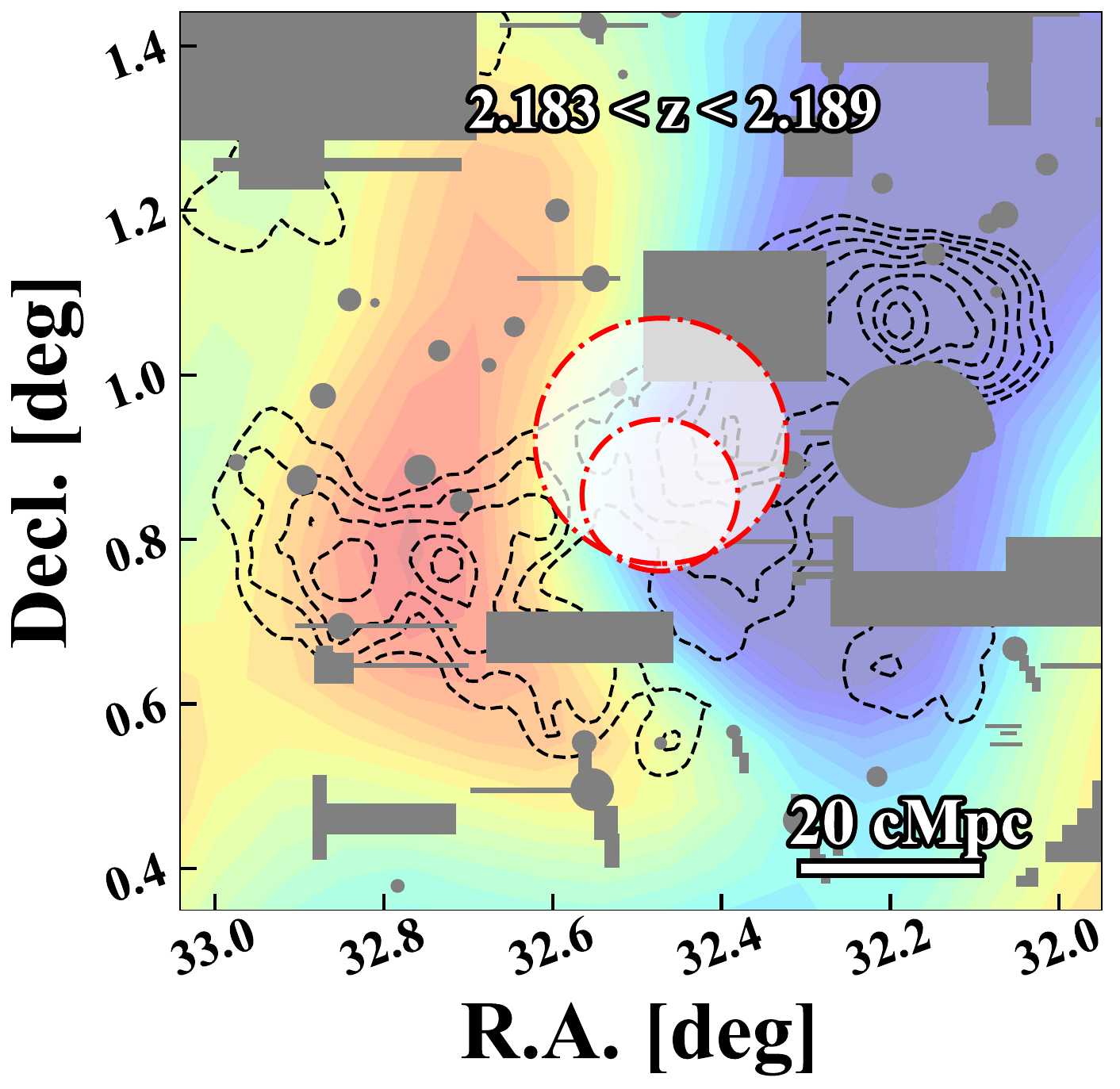}
    \includegraphics[width=0.3\linewidth]{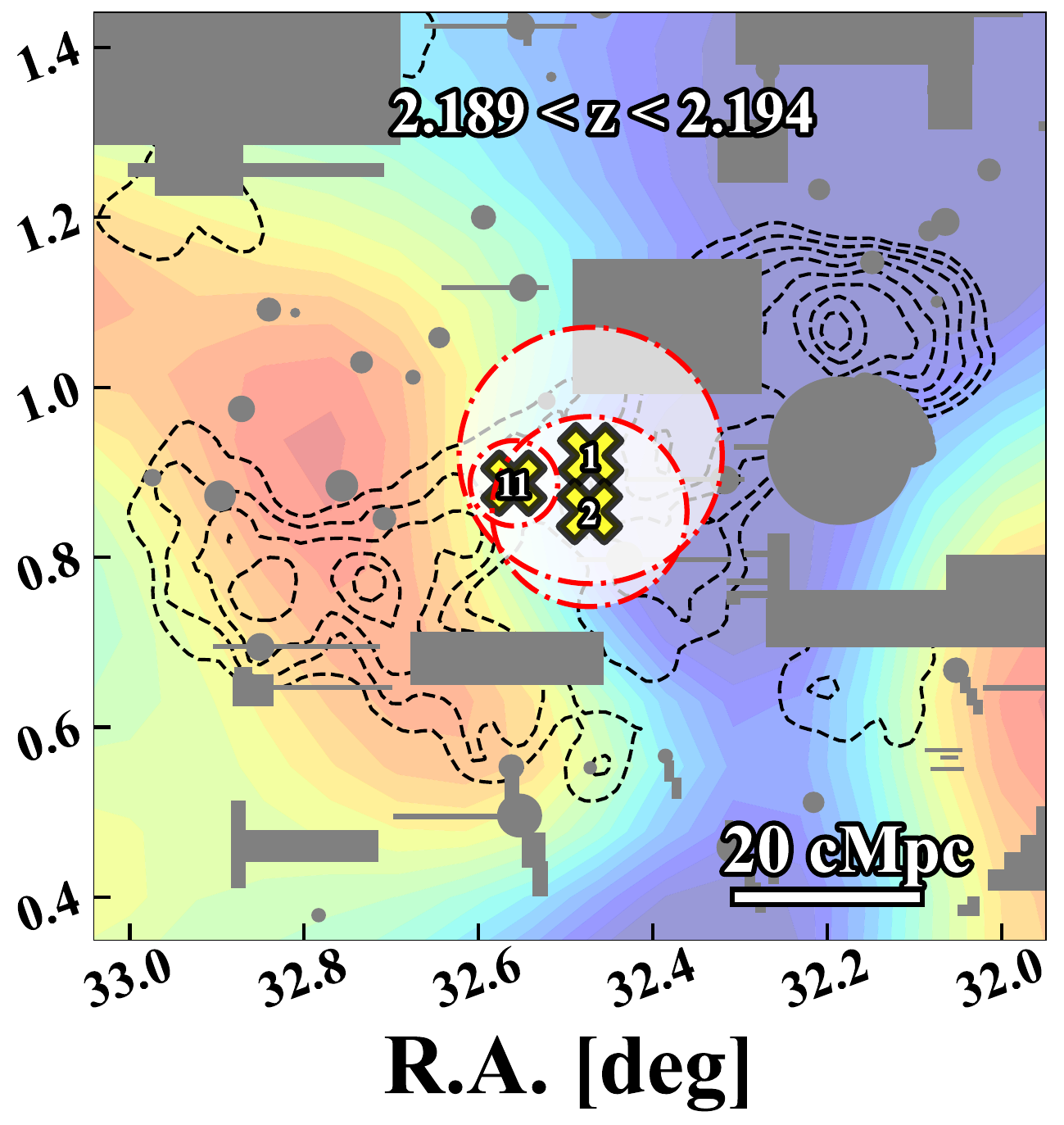}
    \includegraphics[width=0.343\linewidth]{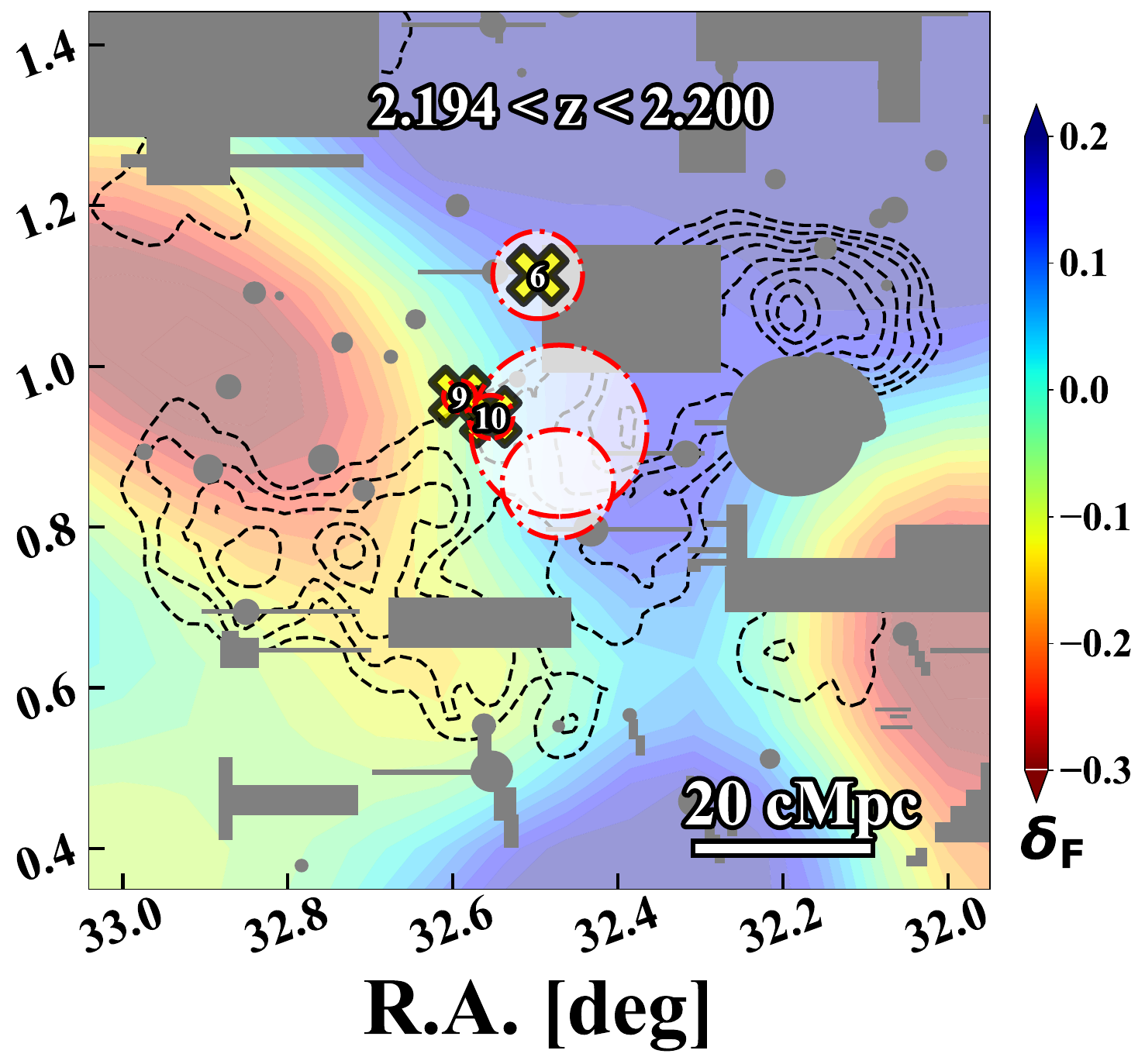}
    \caption{H{\sc i} tomography maps across different redshift slices. 
    The colored contour retains the interpretation from Figure \ref{sec:hi_tomography}, applied to distinct redshift bins. 
    Sequentially from the upper left to the bottom right panel, redshifts increase from $z=2.15$ to $z=2.20$, with intervals indicated in each panel. 
    Yellow crosses mark proximate quasars within the redshift bins, while red dashed-dotted circles represent the photoionizing radius $r_{\rm phot}$. White-shaded areas denote regions where ionizing photons from these luminous SDSS quasars surpass those from the UVB.
    Due to the absence of spectroscopic redshifts for LAEs, their overdensity is consistently depicted as dotted black contours across slices.
    }
    \label{fig:hi_tomo_slice}
\end{figure*}

As indicated in Section \ref{sec:tomo_result_average}, collapsed maps do not suit the examination of individual targets. 
However, they hint at a general trend of proximate quasars congregating near ionizing frontiers. 
Therefore, we delve into individual redshift slices derived directly from the LoSs, constrained by the cell size along the redshift axis at $5.114~h^{-1}{\rm cMpc}$ (or $7.3$ cMpc). 
These slicing maps are displayed in Figure \ref{fig:hi_tomo_slice}, with the colored background contour maintaining the same interpretation as in Figure \ref{fig:hi_trans_map}, but applied to maps at specific redshifts instead of a collapsed overview. 
From the upper left to the lower right panel, the redshift increases from $z=2.150$ to $z=2.200$, with intervals noted in each panel. 
Proximate quasars within each redshift slice are marked as yellow crosses. 
Meanwhile, since LAE candidates are photometrically identified within $z=2.15-2.20$ without precise redshifts, their overdensity is depicted in every panel using black dashed contours.

The IGM H{\sc i} distribution exhibits variation across different redshifts, revealing a 3D structure. 
In every redshift slice, IGM hydrogen in Node-A appears nearly fully ionized, indicating a confidently H{\sc i}-poor environment within this LAE overdensity. 
Conversely, the ionization state of IGM in Node-B varies with redshift. 
Should LAEs within Node-B cluster around $2.17 \lesssim z \lesssim 2.19$, the IGM H{\sc i} overdensity might align with galaxy distributions in a massive structure.  
This pattern also suggests a transition in the IGM ionization stage along the filament. 
The spectroscopic follow-up to determine precise LAE redshifts is crucial for accurately characterizing this vast structure.

The narrative deepens for the proximate quasars. 
The slicing maps reiterate that 10 out of 11 clustering quasars, with the exception of CH-Q03, are situated near the boundary between neutral and ionized regions at every redshift, reinforcing the result that quasars are at or just beyond the ionizing frontiers. 
This result aligns with the $\delta_{\rm F}$ profile described in the collapsed map discussed in Section \ref{sec:tomo_result_average}. 
Thus, it is posited that the clustering quasars may be intricately linked to large-scale IGM ionization processes.  

\section{Discussion}
\label{sec:discussion}

\subsection{The Association between Quasars and IGM {\sc Hi}}
\label{sec:discussion_qso_igm}

\begin{figure*}[hbt!]
    \centering
    \includegraphics[width=\linewidth]{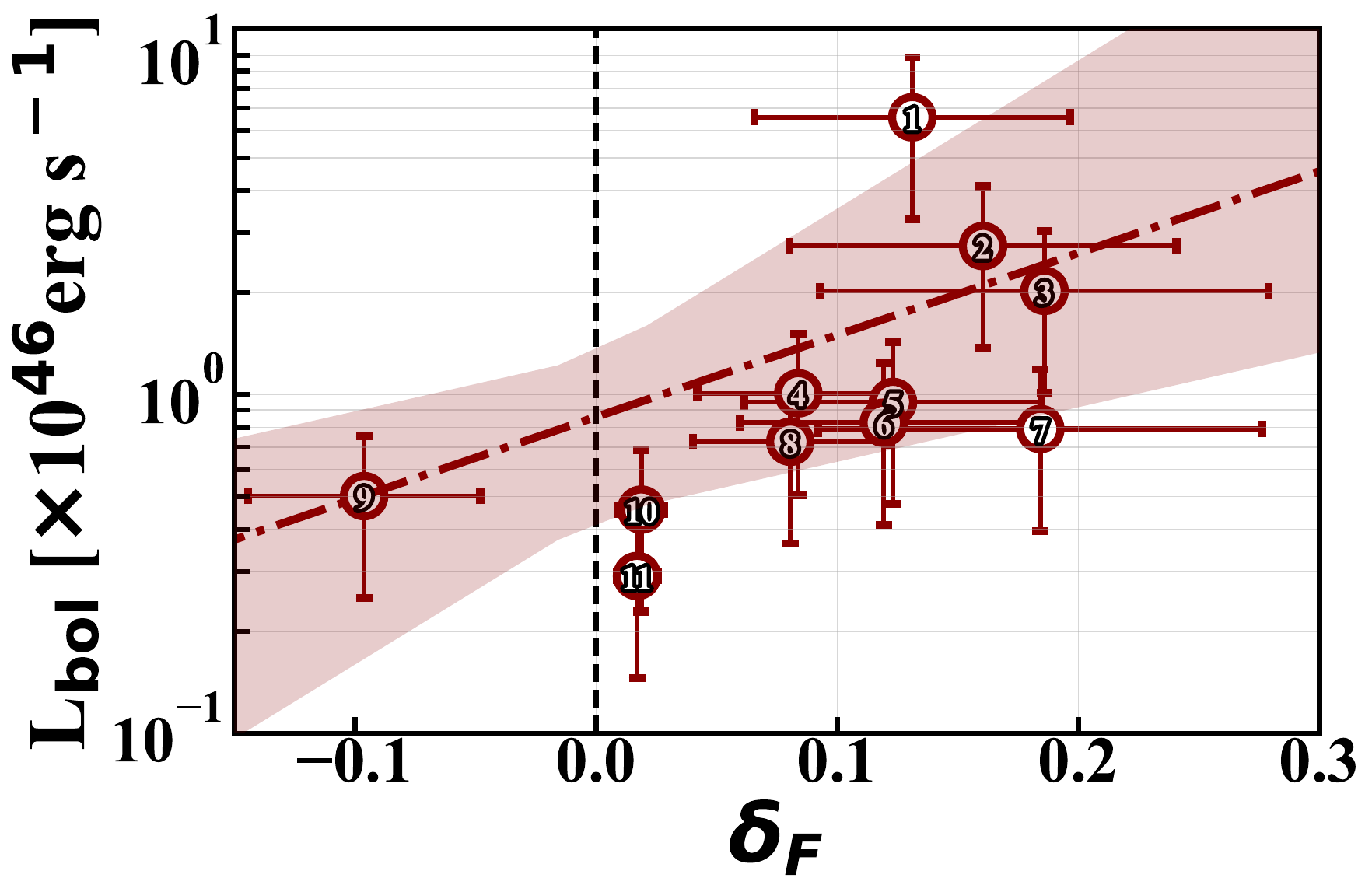}
    \caption{Bolometric luminosity ($L_{\rm bol}$) of proximate quasars plotted against the local Ly$\alpha$ transmission fluctuation, $\delta_F$.  
    Error bars reflect a 50\% uncertainty in converting $L_{\rm 1350}$ to $L_{\rm bol}$. 
    The curve illustrates a power-law relation between $L_{\rm bol}$ and $\delta_{\rm F}$. 
    Confidence intervals are determined through Monte Carlo simulations that vary each data point within a Gaussian distribution using the stated uncertainty; the 16\% and 84\% ranks define the lower and upper bounds, respectively.}
    \label{fig:qso_Lbol_tomo}
\end{figure*}

Our earlier findings indicate that proximate quasars may be associated with IGM H{\sc i} within the underlying structures in 3D space.
Beyond the spatial distribution, we examine the correlation between quasars' bolometric luminosity $L_{\rm bol}$ and $\delta_{\rm F}$ at their respective locations, 
and the $\delta_{\rm F}$ is given by the nearest grid constructed on 15 h$^{-1}$cMpc in Section \ref{sec:hi_tomography}. 
This analysis, depicted in Figure \ref{fig:qso_Lbol_tomo}, reveals a significant positive correlation between $L_{\rm bol}$ and $\delta_{\rm F}$, as evidenced by a Spearman rank test yielding a correlation coefficient $\rho_{\rm S}=0.75$ with a $P$-value of $0.007$. 
A power law fit between $L_{\rm bol}$ and $\delta_{\rm F}$ establishes the relation:
\begin{equation}
    \text{log}\left(L_{\rm bol}\right) = 2.42^{+0.80}_{-1.95} \times \delta_{\rm F} + 45.93^{+0.17}_{-0.31}, 
\end{equation} 
with errors derived from the 16\%--84\% ranks in a Monte Carlo simulation by fluctuating data points with 50\% uncertainties and repeating 1,000 times.
The fitted relation and its uncertainty are illustrated in Figure \ref{fig:qso_Lbol_tomo} by the red dashed-dotted curve and shaded area, respectively. 
This relation further substantiates a strong correlation, indicating that more luminous quasars reside in more transparent IGM environments, thereby suggesting a potential physical linkage between quasars and the IGM on scales exceeding $15~h^{-1}$cMpc.

\begin{figure*}[hbt!]
    \centering
    \includegraphics[width=0.325\linewidth]{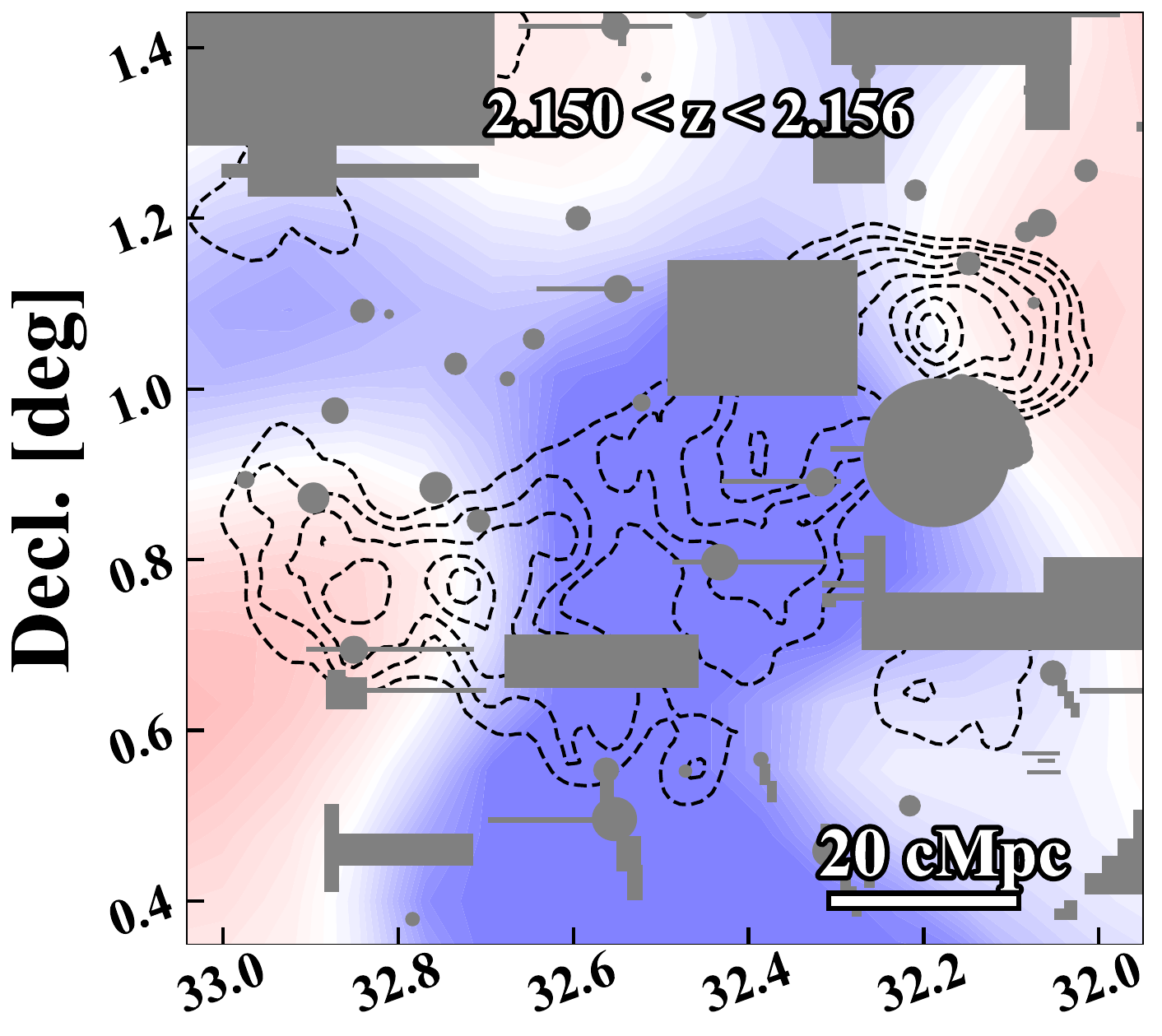}
    \includegraphics[width=0.3\linewidth]{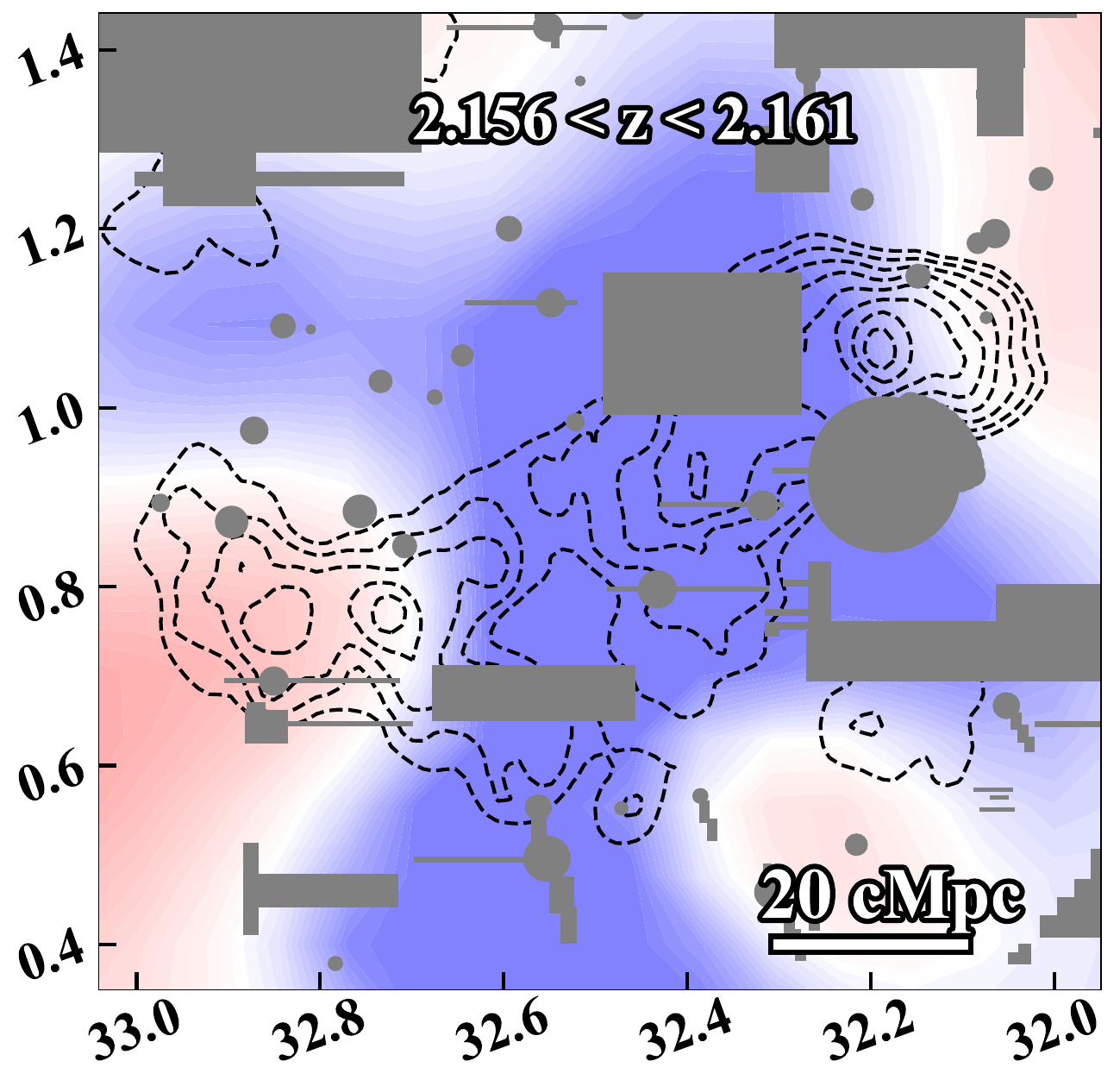}
    \includegraphics[width=0.335\linewidth]{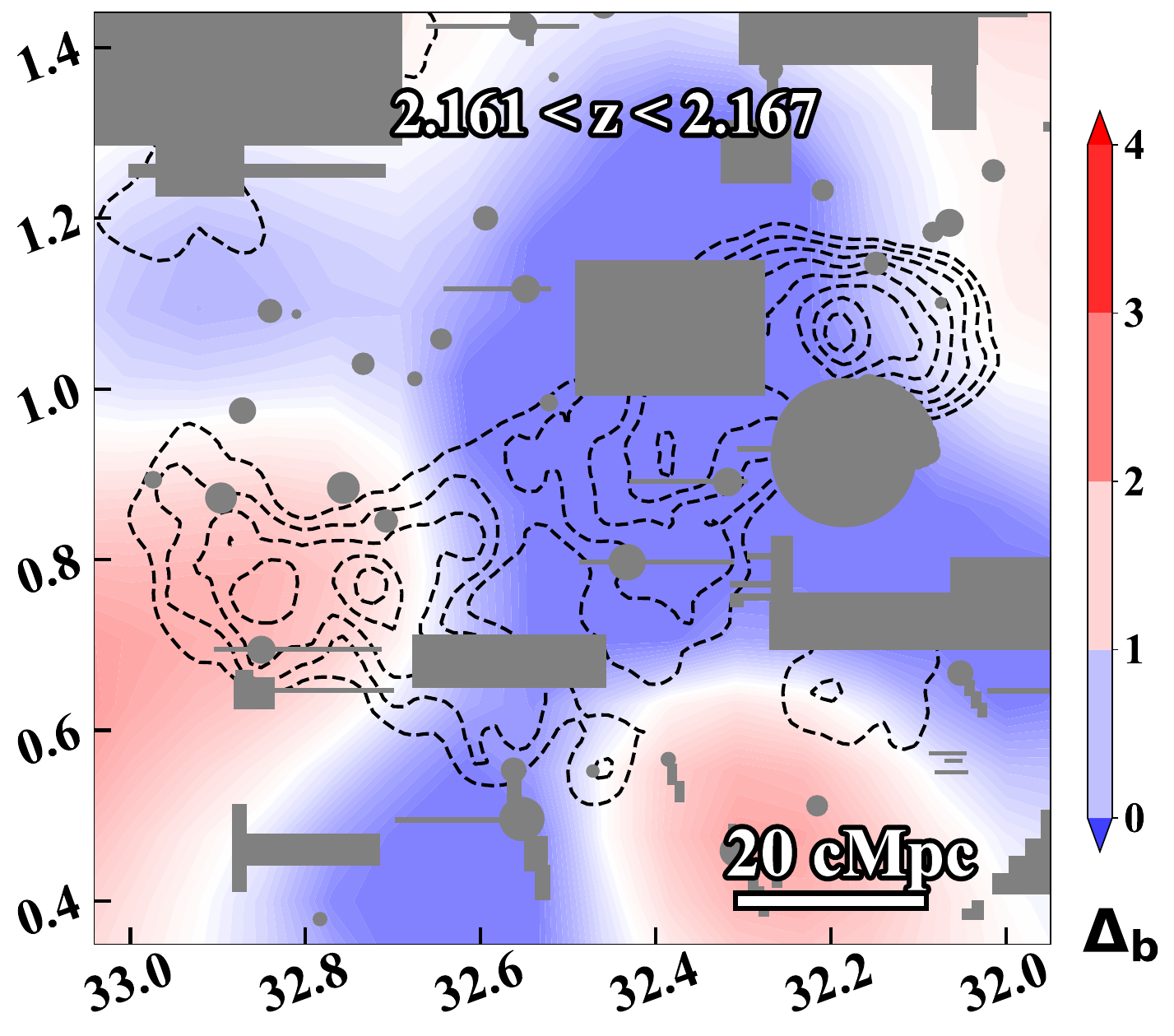}
    \includegraphics[width=0.325\linewidth]{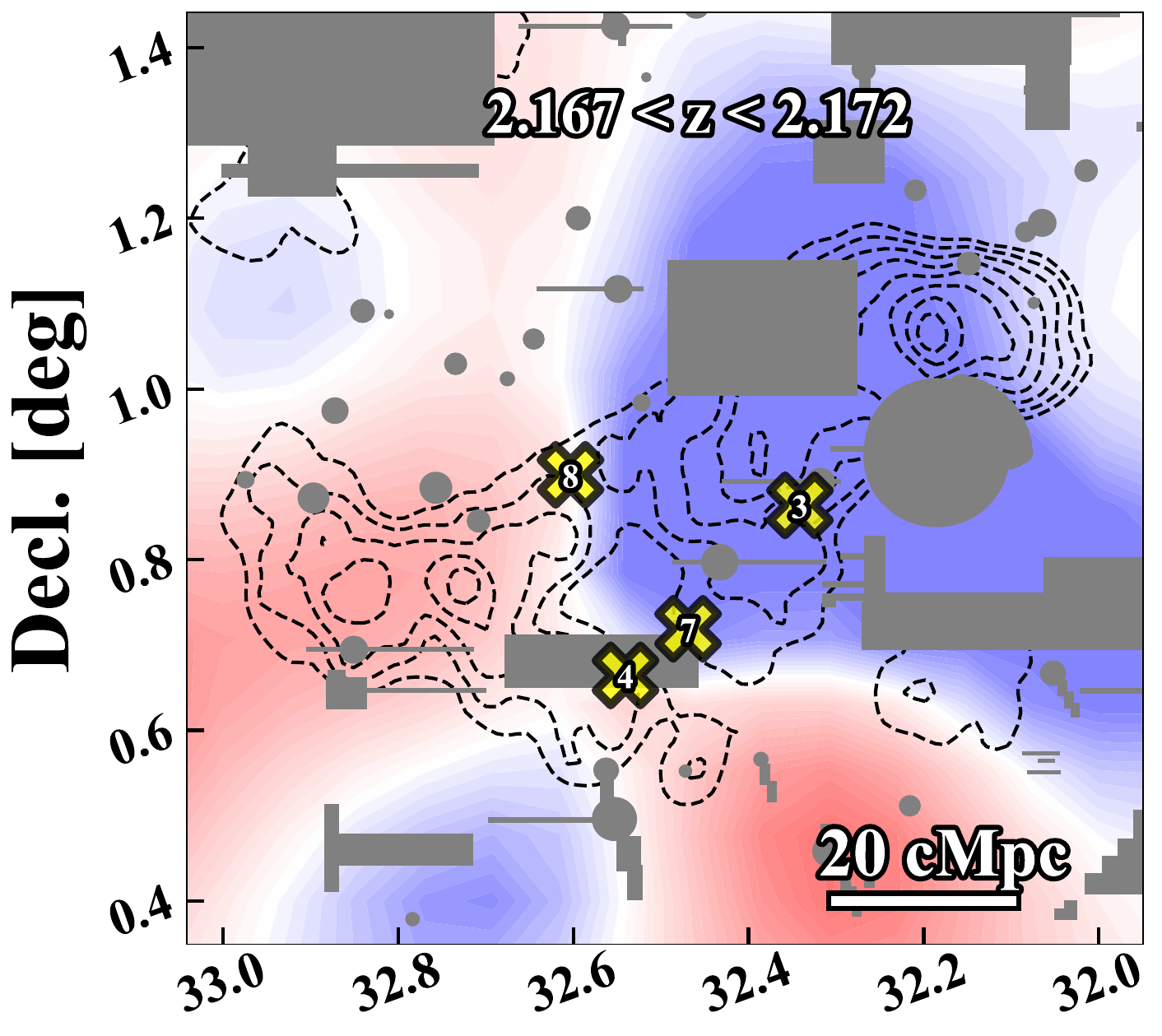}
    \includegraphics[width=0.3\linewidth]{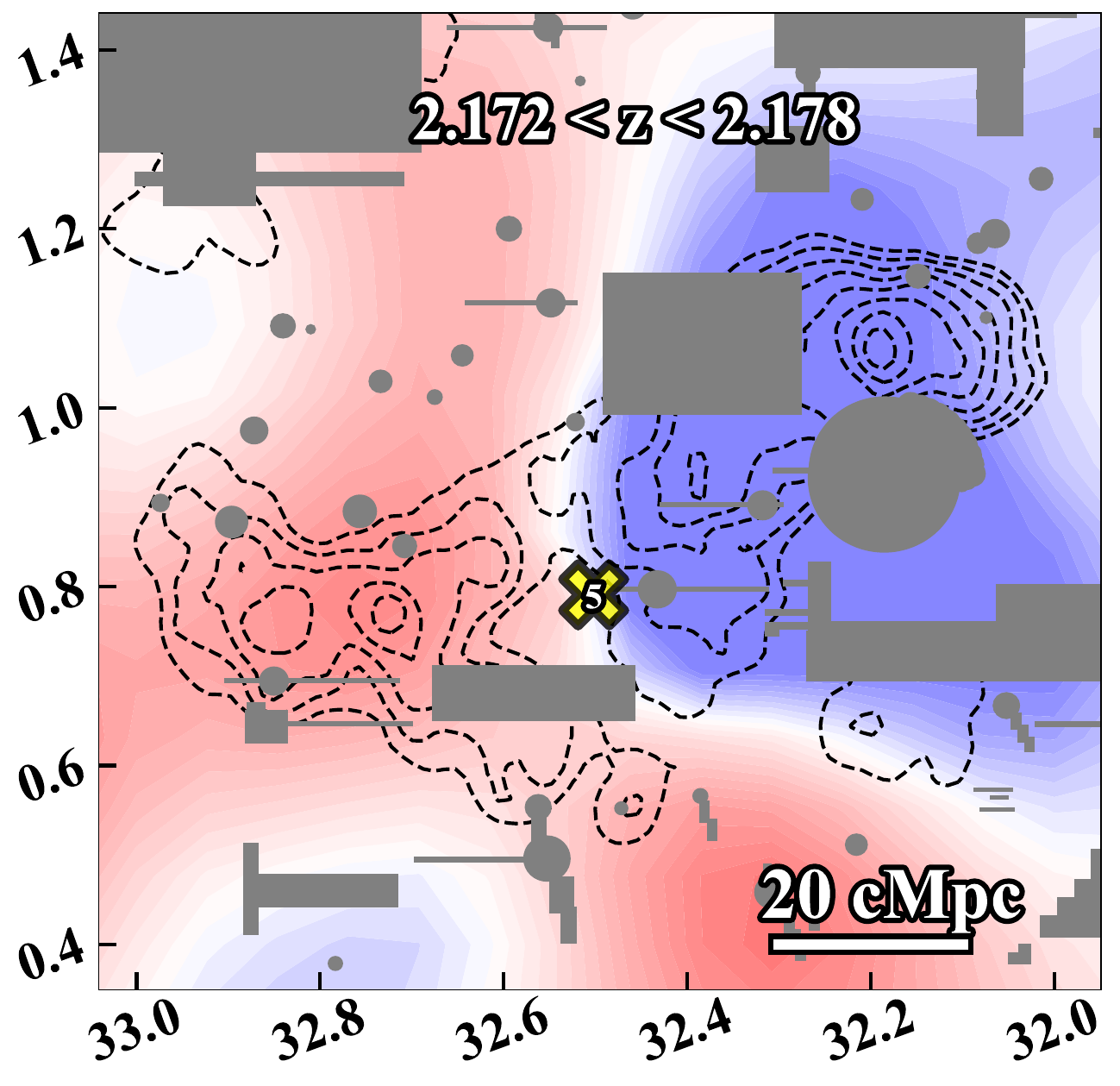}
    \includegraphics[width=0.335\linewidth]{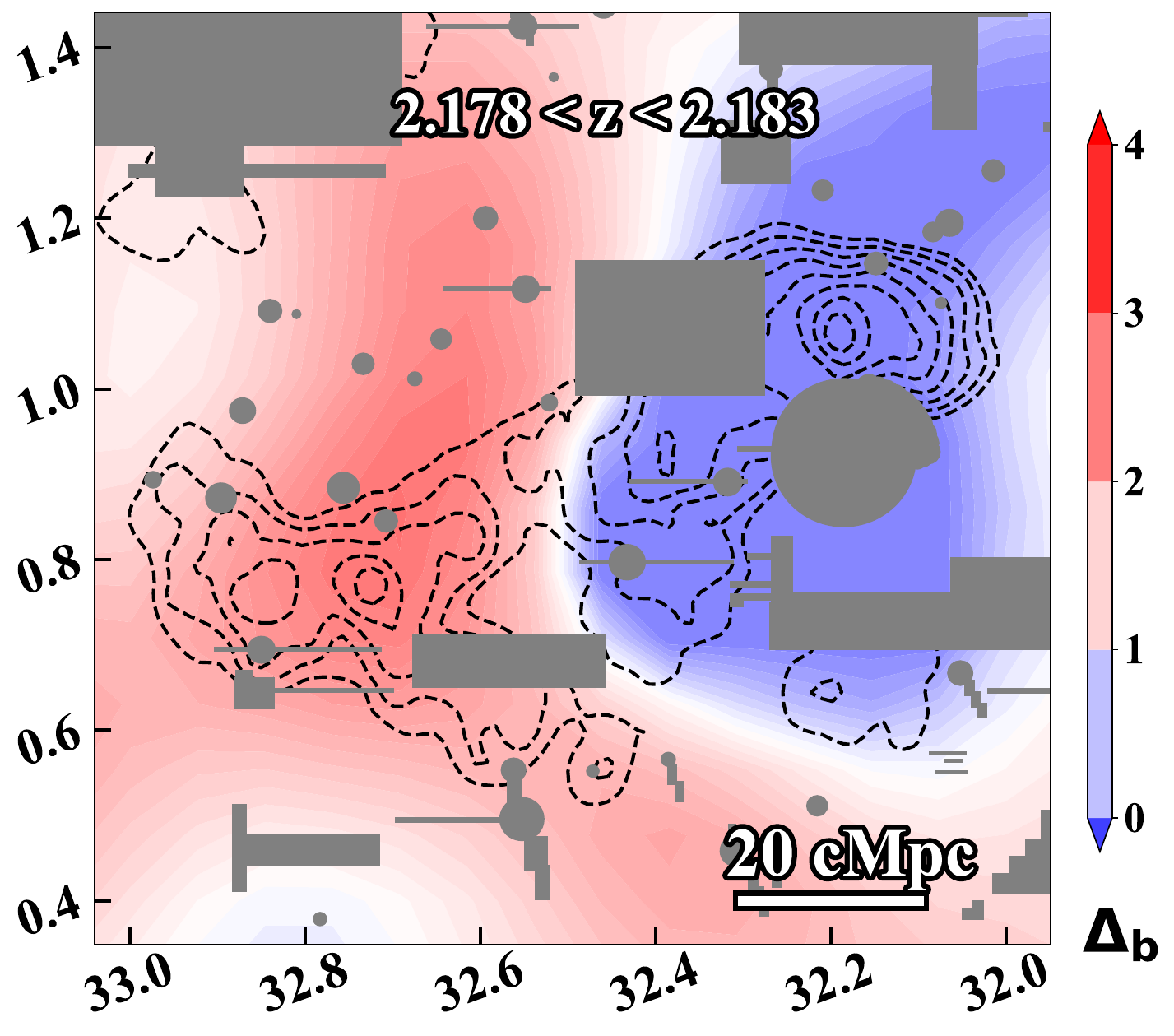}
    \includegraphics[width=0.325\linewidth]{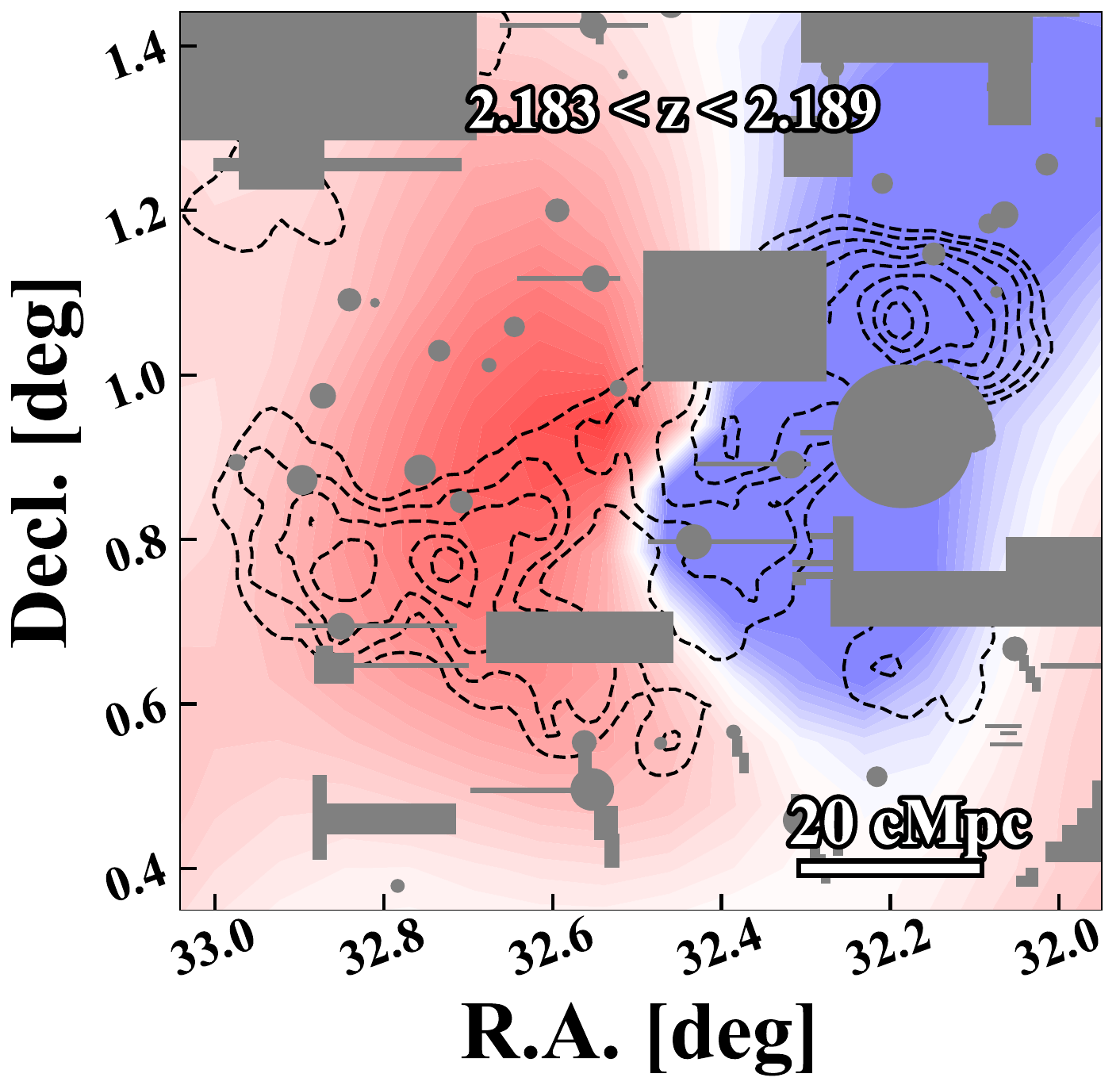}
    \includegraphics[width=0.3\linewidth]{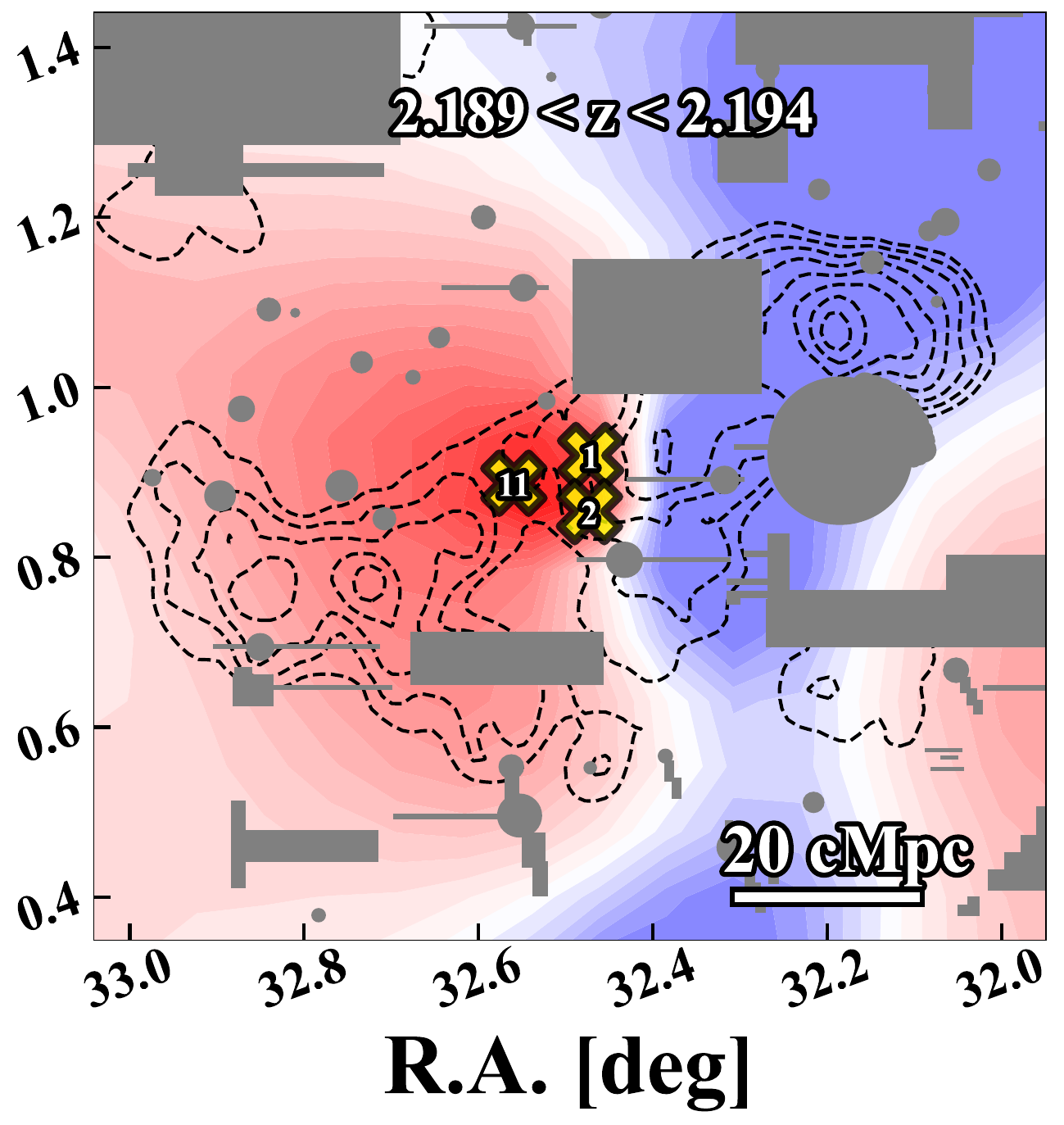}
    \includegraphics[width=0.335\linewidth]{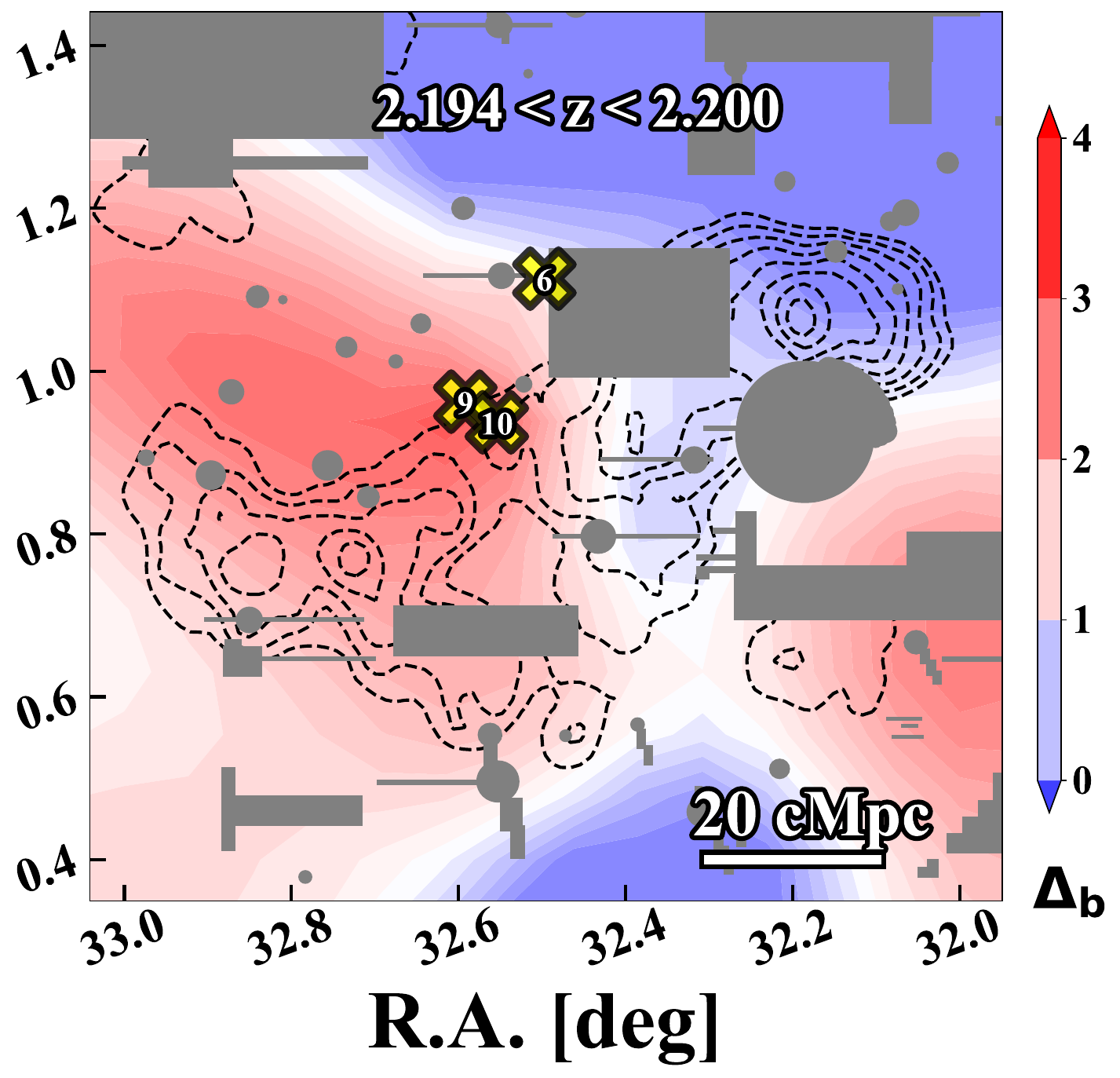}
    \caption{Reconstructed maps of baryon matter fluctuation, $\Delta_b$, across different redshift slices.
    Redder shades signify higher $\Delta_b$ values, representing denser baryonic matter, encompassing both ionized and neutral hydrogen in the IGM. 
    Panel arrangements and symbols used are consistent with those in Figure \ref{fig:hi_tomo_slice}.}
    \label{fig:baryons_map}
\end{figure*}

\begin{figure}[t]
    \centering
    % \hspace*{-1cm}
    \includegraphics[width=\linewidth]{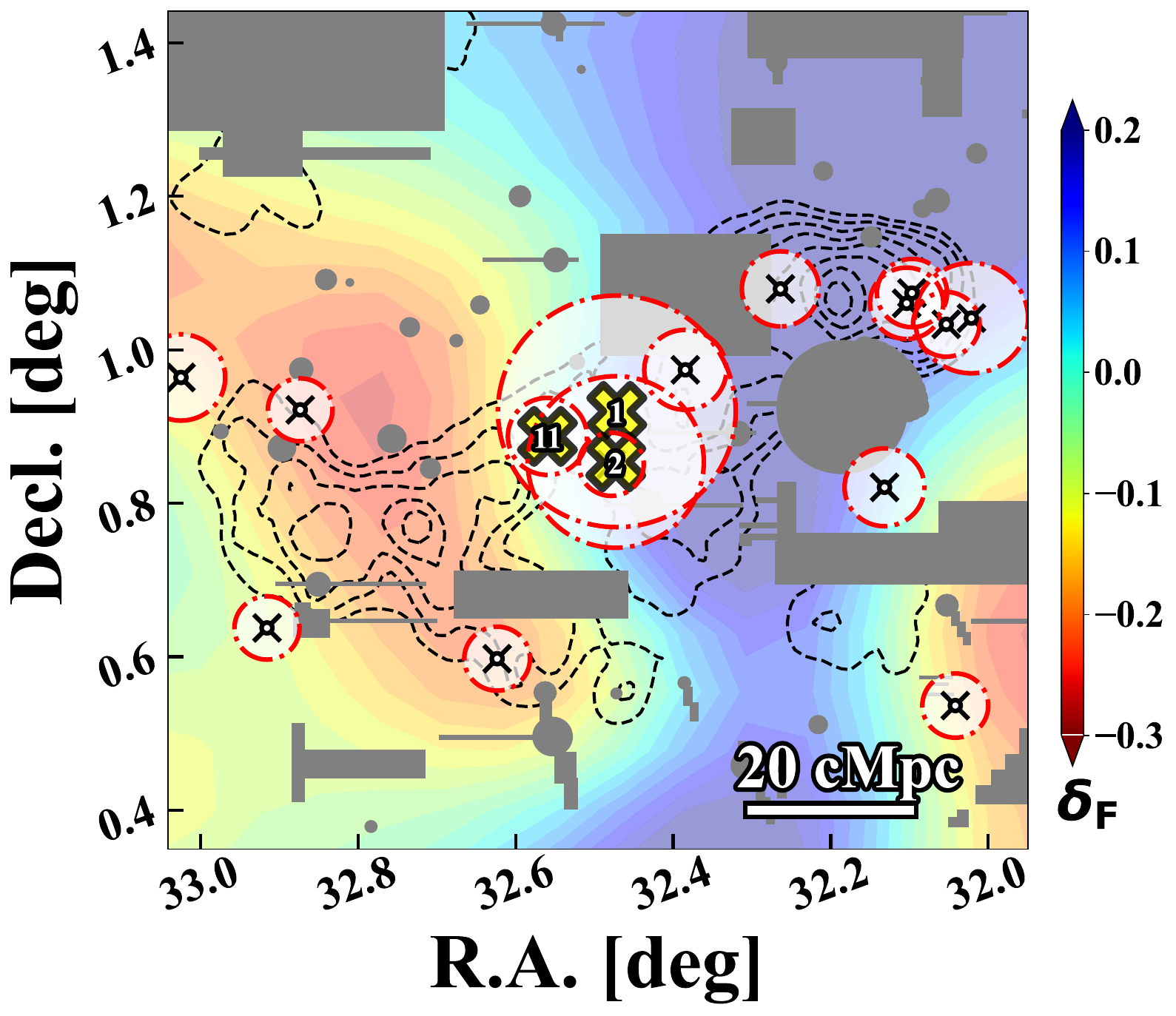}
    \caption{IGM tomography map for $2.189<z<2.194$, akin to Figure \ref{fig:hi_tomo_slice}. 
    Red circles and white spheres denote photoionization sources from both quasars and NUV-LAEs. 
    Quasars are marked as yellow crosses with IDs, while NUV-LAEs are represented by thin black crosses.
    }
    \label{fig:agn_potential_nodea}
\end{figure}

\begin{figure}[hbt!]
    \centering
    \includegraphics[width=\linewidth]{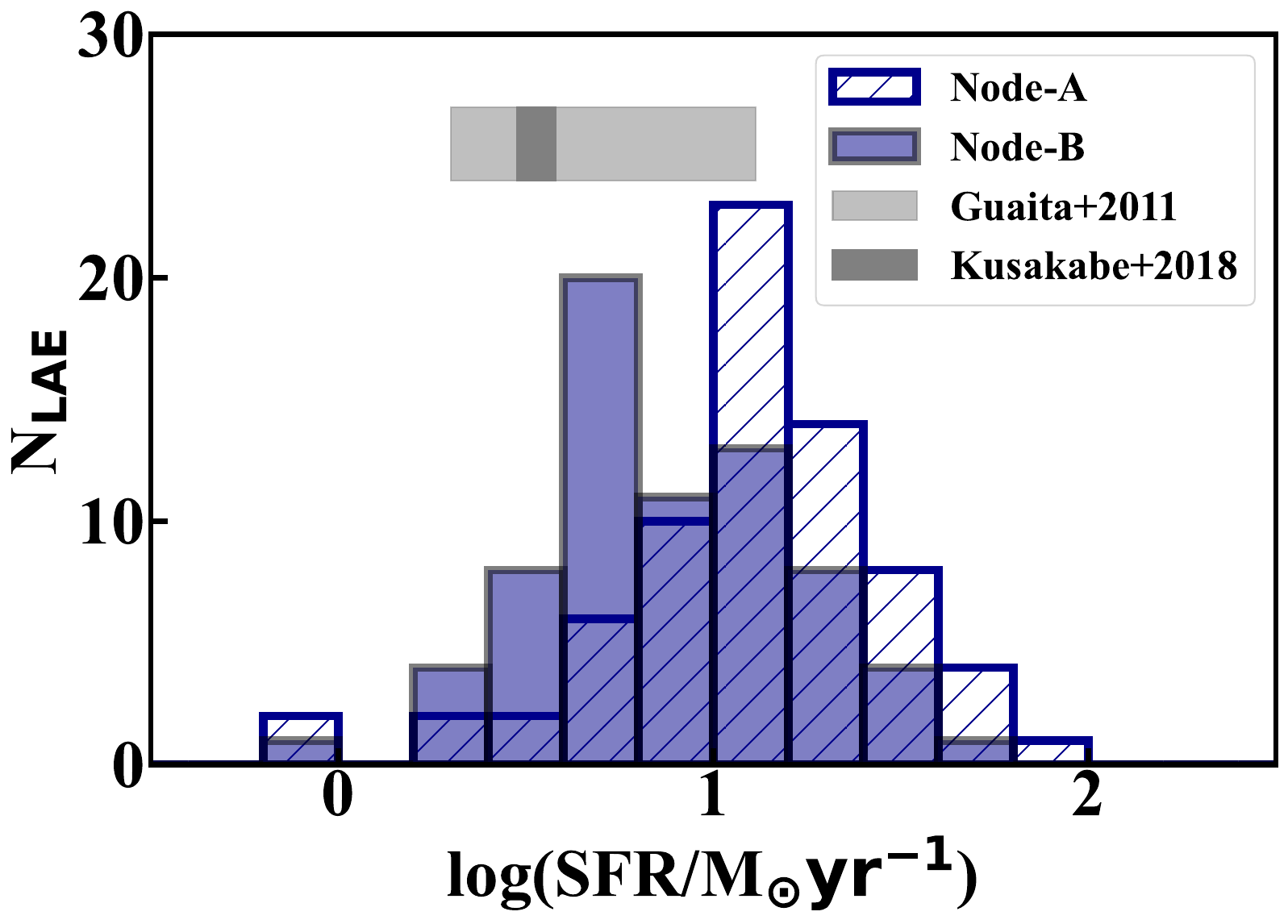}
    \caption{Star formation rate (SFR) distribution of LAEs, with blue hatched and shaded histograms representing member LAEs in Node-A and Node-B, respectively.
    The light and dark gray shaded band suggest 1-sigma results of SFR in the LAEs identified in general fields at $z\approx2.1$ and 2.2 \citep{Guaita+2011, Kusakabe+2018}.
    }
    \label{fig:lae_stellarmass}
\end{figure}

To discern the photoionizing influence of quasars, a comparison between their contributions and that from UV background (UVB) can be insightful, e.g., \citet{Kashikawa+2007}.
We estimate the ionizing photons from quasars in their proximate {\sc Hi} following \citet{ArrigoniBattaia+2015, Pezzulli+2019}.
The photoionization rate for an individual quasar is given by:
\begin{equation}
    \begin{split}
        \Gamma_{\rm HI,~i}^{\rm QSO} = 8.8 \times 10^{-7} \xi_i \left( \frac{r_i}{10~{\rm kpc}}\right)^{-2},
    \end{split}
    \label{eq:phot_xi}
\end{equation}
where the $\Gamma_{\rm HI, i}^{\rm QSO}$ is the photoionization rate of a specific quasar $i$, $\xi_i$ is a factor scaled by this quasar's luminosity and continuum slope, and $r_i$ is the distance between the calculating location and this quasar. 
We note that this $\Gamma_{\rm HI,~i}^{\rm QSO}$ estimate assumes that the quasar radiation is spherically symmetric and their luminosity is time-invariant, which may not be a case in reality, but it still provides instructive ideas. 

Assuming a power-law continuum for wavelengths shorter than the Lyman limit, $\xi_i$ is calculated by rescaling observed $i$-band magnitudes of reference quasars from \citet{Borisova+2016}:
\begin{equation}
    \xi_i = \left(\frac{L_{\rm \nu, LL, i}}{5.2\times10^{31}~{\rm erg~s^{-1}~Hz^{-1}}}\right)\left(\frac{\beta_{\rm QSO, i}+3}{4.7}\right)^{-1},
    \label{eq:xi_luminosity}
\end{equation}
where $L_{\rm \nu, LL, i}$ is the monochromatic continuum luminosity at the Lyman limit (rest-frame 912 \AA), and $\beta_{\rm QSO}$ is the UV continuum slope determined in Section \ref{sec:sdss_qso}. 
By integrating Equations (\ref{eq:phot_xi}) and (\ref{eq:xi_luminosity}), we can identify the radius at which a specified photoionizing rate occurs, defining a photoionizing radius $r_{\rm phot}$ where for distances $r < r_{\rm phot}$: 
\begin{equation}
    \Gamma_{\rm HI,~i}^{\rm QSO} > \Gamma_{\rm HI}^{\rm UVB},
\end{equation}
and $\Gamma_{\rm HI}^{\rm UVB}\sim10^{-12}~{\rm s}^{-1}$ at $z\approx2$ \citep{Haardt+2012}.
Thus, within regions closer than $r_{\rm phot}$ to proximate quasars, the ionizing photons from these luminous SDSS quasars significantly exceed those from the UVB.
Figure \ref{fig:hi_tomo_slice} illustrates the results, with the radius of red dotted circles centered on quasars representing $r_{\rm phot}$. 
Consequently, photoionization within the white-shaded areas is predominantly driven by quasar ionization. 
We note that the shape of these spherical regions can be altered if the emission from quasars is anisotropic. 

Clearly, the most luminous quasars like CH-Q01 and CH-Q02 are capable of influencing the IGM ionization stage up to a distance of 20 cMpc. 
Although the relatively faint quasars may have a more limited impact on their surroundings, their affecting areas can be overlapped to make a coherent contribution to shape the topology of nearby ionizing frontiers.
However, even considering their joint efforts, the $r_{\rm phot}$ of the known eleven quasars likely have little effect on the ionization state of the distant Nodes A and B. 

The quasar and IGM {\sc Hi} spatial distribution, the correlation between quasar $L_{\rm bol}$ and $\delta_F$, and the analysis of $r_{\rm phot}$ collectively indicate that the IGM H{\sc i} on scales 
of tens cMpc 
is possibly modulated by 
the clustering quasars within the \textit{Cosmic Himalayas}. 
Albeit the most luminous quasars may have a more significant impact, as illustrated in the $L_{\rm bol}$ -- $\delta_F$ relation and $r_{\rm phot}$, a single quasar cannot do the whole job to generate a large scale ionized IGM gas structure as observed beyond 20 cMpc. 
Multiple quasars being switched on in a relatively recent period, at least no earlier than the IGM {\sc Hi} recombination timescale, are essential to produce this peculiar ionizing structure.

\subsection{The Potential  
Baryonic Matter Distribution}
\label{sec:discussion_baryons}

We have noted that the quasar overdensity does not coincide with the peaks of LAE density or the peaks of IGM H{\sc i} absorption/transmission within the J0210 field. 
However, the distribution of total baryonic matter, including ionized hydrogen, has yet to be clarified. 
Utilizing the IGM H{\sc i} tomography map, it is possible to reconstruct the distribution of baryonic matter by accounting for the photoionization effects attributed to quasars.

In the IGM, the {\sc Hi} Ly$\alpha$ optical depth, $\tau_{\rm Ly\alpha}$, is related to the {\sc Hi} column density, N$_{\rm HI}$, through the equation:  
\begin{equation}
    \tau_{\rm Ly\alpha} = \sigma_{\rm Ly\alpha}{\rm N}_{\rm HI},
    \label{eq:tau_nHI}
\end{equation}
where $\sigma_{\rm Ly\alpha}$ is the \lya~cross section. 
The N$_{\rm HI}$ can be determined using:
\begin{equation}
    \begin{split}
        {\rm N}_{\rm HI} & = l_{\rm Ly\alpha} n_{\rm HI} \\
        & = l_{\rm Ly\alpha} \left<n_{\rm H}\right> \Delta_b x_{\rm HI},
    \end{split}
    \label{eq:nHI_xHI}
\end{equation}
in which $l_{\rm Ly\alpha}$ represents the line-of-sight length of the medium that \lya~photons pass through, $n_{\rm HI}$ and $\left<n_{\rm H}\right>$ are the local number density of neutral hydrogen and cosmic mean hydrogen, $\Delta_b$ signifies the baryon fluctuation, and $x_{\rm HI}$ is the {\sc Hi} fraction. 
As indicated in Section \ref{sec:hi_tomo_method}, at $z\sim2.2$, cosmic H{\sc i} is predominantly ionized with $\langle F(z) \rangle_{\rm cos} \sim 0.84$. Consequently, under conditions where photoionization reaches equilibrium, it is reasonable to consider case A (optically thin) hydrogen recombination.
Assuming the equilibrium can be established on a very short time-scale, $x_{\rm HI}$ can be approximated following \citet{Meiksin+2009}:
\begin{equation}
    x_{\rm HI} = \frac{\Gamma_{\rm rec}}{\Gamma_{\rm HI}^{\rm tot}} = \frac{\left<n_{\rm H}\right> \Delta_b \alpha(T)}{\Gamma_{\rm HI}^{\rm tot}},
    \label{eq:xHI_phot}
\end{equation}
where $\Gamma_{\rm rec}$ and $\Gamma_{\rm HI}^{\rm tot}$ represent the hydrogen recombination rate and the total {\sc Hi} photoionization rate, respectively. 
The $\alpha(T)$ is the Case A hydrogen recombination coefficient at temperature $T$, and typically, $T\approx10^4$ K in IGM.  
Merging Equations (\ref{eq:tau_nHI}), (\ref{eq:nHI_xHI}), and (\ref{eq:xHI_phot}) leads to:
\begin{equation}
    \tau_{\rm Ly\alpha}=\frac{\sigma_{\rm Ly\alpha} l_{\rm Ly\alpha} \left<n_{\rm H}\right>^2 \Delta_b^2 \alpha(T)}{\Gamma_{\rm HI}^{\rm tot}}.
    \label{eq:tau_phot}
\end{equation}
On the other hand, the hydrogen recombination rate can be expressed as \citep{Storey+1995}:
\begin{equation}
\label{eq:recomb_temp}
    \alpha_A(T) = 4.2 \times 10^{-13} \left(T/10^4 K\right)^{-0.7}~{\rm cm}^3~{\rm s}^{-1},
\end{equation}
while the temperature--density relation in IGM follows:
\begin{equation}
\label{eq:temp_density}
    T=T_0 \Delta_b^{\gamma-1},
\end{equation}
where $\gamma\sim1.6$ at $z=2$ \citep{Hui+1997}.
Combining Equations \ref{eq:recomb_temp} and \ref{eq:temp_density}, we have:
\begin{equation}
    \alpha_A(T) \propto \Delta_b^{-0.42}
\end{equation}
Hence, Equation (\ref{eq:tau_phot}) implies:   
\begin{equation}
\label{eq:tau_deltab}
    \begin{split}
        \tau_{\rm Ly\alpha} & \propto \frac{\Delta_b^{1.58}}{\Gamma_{\rm HI}^{\rm tot}} \\
        & \propto \frac{\Delta_b^{1.58}}{B \times \Gamma_{\rm HI}^{\rm UVB}},
    \end{split}
\end{equation}
where the boost factor $B$ incorporates the quasar contribution:
\begin{equation}
    B=\left(\Gamma_{\rm HI}^{\rm UVB} + \Gamma_{\rm HI}^{\rm QSO}\right)/\Gamma_{\rm HI}^{\rm UVB}.
\end{equation}
The $\Gamma_{\rm HI}^{\rm UVB}$ is the UV background photoionization rate and $\Gamma_{\rm HI}^{\rm QSO}$ denotes the photoionization rate from proximate quasars.
Meanwhile, by definition, the {\sc Hi} optical depth is 
\begin{equation}
\label{eq:tau_definition}
    \begin{split}
        \tau_{\rm Ly\alpha} & = - {\rm ln}\left(F\right) \\
        & = - {\rm ln}\left[ \left(1+\delta_F \right) \left<F\right>_{\rm cos}\right],
    \end{split}
\end{equation}
with $F$, $\left<F\right>_{\rm cos}$, and $\delta_F$ representing the local Ly$\alpha$ transmission, cosmic mean Ly$\alpha$ transmission, and local Ly$\alpha$ transmission fluctuation, respectively.
By connecting Equations (\ref{eq:tau_deltab}) and (\ref{eq:tau_definition}), a proportional relation is established:
\begin{equation}
    \Delta_b \propto \left\{ - \Gamma_{\rm HI}^{\rm tot} \times {\rm ln}\left[ \left( 1+ \delta_F\right) \left<F\right>_{\rm cos} \right]\right\}^{1/1.58}.
\end{equation}
When $\delta_F=0$ and $\Gamma_{\rm HI}^{\rm tot}=\Gamma_{\rm HI}^{\rm UVB}$, there should be $\Delta_b=\rho_b/\left<\rho_b\right>=1$, with $\rho_b$ and $\left<\rho_b\right>$ denoting the local density and cosmic mean density of baryons.
Hence, we get:
\begin{equation}
    \begin{split}
        \Delta_b & = \left\{ 
        \frac{\Gamma_{\rm HI}^{\rm tot} \times {\rm ln}\left[ \left( 1+ \delta_F\right) \left<F\right>_{\rm cos} \right]}
        {\Gamma_{\rm HI}^{\rm UVB} \times {\rm ln}\left<F\right>_{\rm cos}}
        \right\}^{1/1.58} \\
        & = \left\{ B \times \left[1 + \frac{{\rm ln}\left( 1 + \delta_F\right)}{{\rm ln}\left<F\right>_{\rm cos}}\right]
        \right\}^{1/1.58},
    \end{split}
\end{equation}
revealing that the baryonic fluctuation $\Delta_b$ can be deduced from Ly$\alpha$ transmission fluctuation $\delta_F$ and the boost factor $B$ due to quasar proximity.
To compute $B$ across the IGM tomography map's grids, the collective photoionization rate from all eleven quasars is derived from Equation (\ref{eq:phot_xi}):
\begin{equation}
    \begin{split}
        \Gamma_{\rm HI}^{\rm QSO} 
        & = \sum_i^{11} \Gamma_{\rm HI, i}^{\rm QSO} \\
        & = \sum_i^{11} 8.8 \times 10^{-7} \xi_i \left( \frac{r_i}{10~{\rm kpc}}\right)^{-2}.
    \end{split}
    \label{eq:phot_sum}
\end{equation}

The 3D tomography map depicting the baryonic matter fluctuation, $\Delta_b$, is presented in Figure \ref{fig:baryons_map}, with a redder color indicating denser matter. 
This aligns with the discussions in Section \ref{sec:discussion_qso_igm} about quasars of relatively low luminosity having minimal impact on the IGM ionizing stage on large scales. 
The reconstructed $\Delta_b$ distribution does not exhibit a significant deviation from $\delta_F$. However, the predominant photoionization influence of CH-Q01 and CH-Q02 results in a substantial elevation of $\Delta_b$ around these luminous quasars compared to H{\sc i}. 
This emphasizes the presence of a matter-density peak surrounding the quasars—an effect that cannot be directly inferred from H{\sc i} tomography alone without considering ionization effects. Moreover, the inferred matter-density distribution differs from the observed LAE distribution, indicating a potential bias in galaxy populations, as discussed in Section \ref{sec:discussion_qso_trigger}.

Although the reconstructed $\Delta_b$ is referred to as the baryon fluctuation, it is derived solely from the {\sc Hi} distribution, which may introduce a bias relative to {\sc Hii}. Additionally, the reconstruction only accounts for the ionization effects of known SDSS/eBOSS quasars, potentially overlooking other ionizing sources. A particularly striking feature is the sharp decline in $\Delta_b$ between the quasar overdensity and Node-A (Figure \ref{fig:baryons_map}), which appears unphysical and suggests the presence of missing ionization sources (as discussed in Section \ref{sec:discussion_nodea_ionize}). This highlights the need to critically assess the strong assumptions in Section 5.2 to refine our understanding of the physical processes shaping the \textit{Cosmic Himalayas}.

\subsection{The Lack of IGM {\sc Hi} in Node-A}
\label{sec:discussion_nodea_ionize}

The H{\sc i} structure around LAE Node-A remains more puzzling, as shown in Figures \ref{fig:hi_trans_map} and \ref{fig:hi_tomo_slice}. 
The $\delta_F$ in Node-A remains consistently around 0.2, indicating an anomalously transparent IGM at $z\approx2.2$ despite a high LAE density. Given that the nearest SDSS/eBOSS quasars are at least 25 cMpc away, their photoionization impact on Node-A is negligible, as quantified by $r_{\rm phot}$. 
More notably, the $\Delta_b$ map (Figure \ref{fig:baryons_map}) exhibits a sharp and physically implausible decline in density between the quasar overdensity and Node-A within $2.189<z<2.194$. This abrupt drop reinforces the hypothesis that additional ionization sources, unaccounted for in our calculations, may be responsible for the observed IGM transparency.

One possibility is that the optically obscured AGNs responsible for the ionizing photon budget are overlooked in the SDSS/eBOSS quasar selection process.
This brings attention to NUV-LAEs, considered potential AGN candidates due to their NUV excess in the Lyman break regime. 
Assuming, in the most extreme scenario, that all NUV-LAEs are indeed AGNs and cluster around $2.189<z<2.194$, we can calculate the photoionization radius $r_{\rm phot}^{\rm NUV-LAE}$ for these sources using a methodology akin to that discussed in Section \ref{sec:discussion_qso_igm}. 
The $\xi$ for NUV-LAEs were acquired through extrapolating an equivalent $i$-band from the NUV band assuming a power-law continuum with typical $\beta_{\rm QSO}$. 
The $r_{\rm phot}^{\rm NUV-LAE}$, represented by the radius of red dashed-dotted circles centered on black crosses, is illustrated in Figure \ref{fig:agn_potential_nodea}. This analysis suggests that, even in the most extreme case, the NUV-LAEs alone are insufficient to account for the fully ionized IGM around Node-A.

Another potential contributor could be 
excess star-formation activity in 
star-forming galaxies within these regions, 
if we assume the escape fraction of ionizing photons can be comparable in the two structures. 
To explore this, we assess the UV star formation rate (SFR) of member LAEs in Node-A and Node-B, accounting for 71 and 74 member LAEs, respectively, after excluding the NUV-LAEs.
We derive the UV SFR based on the scaling relation given by \citet{Madau+1998}:
\begin{equation}
    {\rm SFR}/{\rm M_{\odot}~yr^{-1}} = \frac{L_{\rm 1500}/{\rm erg}~{\rm s}^{-1}~{\rm Hz}^{-1}}{8.0 \times 10^{27}},
\end{equation}
where $L_{\rm 1500}$ represents the UV flux at $\lambda_0=1,500$~\AA.
A correction factor of $0.08$ is applied to the HSC $g$-band magnitude to account for the wavelength offset, assuming a typical UV slope of LAEs as $\beta=-1.6$ \citep{Kusakabe+2018}.
The SFR distributions for LAEs are depicted in Figure \ref{fig:lae_stellarmass}, showing a noticeable difference, as confirmed by a two-sample K-S test with a $p$-value $<10^{-4}$.
Compared to the LAEs in general fields, Node-B members share a SFR distribution almost similar to \citet{Guaita+2011}, and slightly higher than \citet{Kusakabe+2018}, but Node-A members have systemically higher star formation activity than both. 
As a result, the ionizing photon production from star-forming activities in Node-A is systematically higher than that in Node-B by approximately 0.3 dex. 
This significant amount suggests that the elevated SFR around Node-A could partly account for the ionized state of the IGM in its vicinity.

Furthermore, the combined effects of obscured AGNs and star-forming galaxies might collaboratively enhance the photoionization within Node-A. 
The potential for a more refined IGM tomography, aiming for a resolution closer to $\sim 3$ cMpc in future studies with the upcoming Subaru Prime Focus Spectrograph (PFS), promises to delineate their contributions more precisely.

\subsection{Possible Quasar Triggers in The Densest Quasar Group}
\label{sec:discussion_qso_trigger}

Quasars are often considered the aftermath of gas-rich 
major mergers \citep{Hopkins+2006}.
This hypothesis has garnered observational support at $z<1$ based on the statistical analysis based on galaxy morphology \citep[e.g.,][]{Goulding+2018}.
Therefore, active quasar activities can be a proxy of rich environments with galaxy overdensities.
Nonetheless, the quasars within the \textit{Cosmic Himalayas} deviate from this pattern.
They do not coincide with the density peaks of LAEs. 
Instead, they appear prominently positioned at the midpoint of a large-scale filament. 
This fact complicates the conventional understanding of quasars' origins in major mergers, suggesting a more nuanced interplay between quasar activity and its surrounding cosmic structures.

To reconcile the observed quasars of the \textit{Cosmic Himalayas} with the standard model that links quasar activity to major mergers, a picture is favored that the quasars are hosted by massive galaxies and they are also surrounded by massive galaxies.
While it is widely known that massive galaxies with $M_*\gtrsim10~M_{\odot}$ tend to have weak Ly$\alpha$ emission, e.g., EW$_0 < 20$ \AA~\citep[e.g.,][]{Oyarzun+2016}, it is puzzling why the massive galaxies are segregated from the low-mass LAEs around the clustered quasars.
Here, we consider several potential explanations that can help resolve the apparent discrepancy.

First, major mergers, capable of triggering quasar activity, are also expected to generate significant amounts of dust, rendering the quasar's environment highly dusty by kinetic feedback from jets \citep{Hopkins+2008}. 
This could hinder the detection of resonant Ly$\alpha$ photons from galaxies, leading to an observational bias where many LAEs proximate to quasars might be overlooked in NB imaging \citep[e.g.,][]{Shimakawa+2017b, Ito+2021}.
It is noteworthy that a clear discrepancy between dusty star-forming galaxies and the more typical star-forming galaxies with H$\alpha$ emission show out in two massive protoclusters also traced by multiple quasars \citep{ZhangYH+2022}.
Alternatively, the influence of quasars on their surroundings might manifest as strong radiative feedback mechanisms that heat the cold gas in their vicinity, thereby inhibiting the recent formation of galaxies. 
It is worth noting that while \citet{Dong+2023, Dong+2024} have demonstrated in simulations that radiative feedback can ionize the IGM on scales exceeding $10~h^{-1}$ Mpc, more direct observational evidence of AGN jets influencing the cosmic web on $\sim5$ Mpc scales has recently been reported by \citet{Oei+2024} through radio observations.
Therefore, it is possible that AGN feedback plays a role in regulating the presence and detectability of young LAEs near quasars, though further investigation is needed to fully assess its impact.

If these scenarios hold, the actual galaxy overdensity within the \textit{Cosmic Himalayas} could be significantly higher than what is currently observed. 
Subsequently, the estimated present-day halo mass of $M_h(z=0)\sim10^{14.8}~{\rm M}_{\odot}$ for Node-A, derived from the distribution of detectable LAEs, would merely represent a lower limit of the true galaxy density and mass within this region.

Another possibility of a collision between galaxy (proto-)clusters,  
extending beyond the conventional galaxy merger origin, emerges from the structure delineated by LAEs. 
The two nodes seem to be linked by a cosmic filament, with quasars positioned perpendicularly to this filament both in the projected plane and in redshift space, as illustrated in Figure \ref{fig:hi_trans_map}. 
This configuration bears resemblance to morphological analogs in the local universe, e.g., the Bullet Cluster (1E 0657-558) at $z=0.296$ \citep{Tucker+1998, Clowe+2004} and the Sausage and Toothbrush \citep[CIZA J2242.8+5301 and 1RXS J0603.3+4213;][]{Stroe+2015}.
These structures represent the aftermath of a collision between two galaxy clusters, leading to intensive interaction of their intracluster gases. 
Consequently, the gas, having lost significant momentum, was separated from the dark matter halos, manifesting as strong X-ray emission in shock regions. 
Notably, an elevated proportion of X-ray luminous AGNs has been observed within the Bullet Cluster  
\citep{Puccetti+2020}, hinting at mechanisms that might facilitate increased gas inflow to activate AGNs during such galactic cluster collisions.

It is crucial to recognize the disparity in spatial scale between the Bullet Cluster (approximately $2~{\rm cMpc}$) and the LAE filament in J0210 (around $100~{\rm cMpc}$). 
Nonetheless, given the potential size evolution of protoclusters, the LAE filament with its two nodes could represent a progenitor of structures similar to the Bullet Cluster at $z=2.2$ \citep{Chiang+2013}. 
To validate this hypothesis, further investigations, such as deep X-ray observations of the diffuse intergalactic gas and detailed kinematic studies, are essential.

\section{Summary}

In this paper, we report the discovery of an extraordinary cosmological structure in the J0210 field, the \textit{Cosmic Himalayas}, traced by both the presence of a significant quasar concentration and the distribution of LAEs indicative of large-scale cosmic filaments. 
Utilizing data from SDSS/eBOSS for quasar identification and 3D IGM tomography and Subaru/HSC narrowband imaging for LAE selection, we present a comprehensive analysis revealing the spatial distribution of quasars, LAEs, and the IGM \textsc{Hi}.
Important results include:
\begin{itemize}
\item The discovery of an unprecedented concentration of eleven type-1 AGNs at $z \approx 2.16-2.20$ within a compact $(40~\mathrm{cMpc})^3$ region, marking the densest quasar overdensities $\delta_{\rm QSO}\approx30$ with $16.9\sigma$ significance at $z>2$, situated at the midpoint of a $100$ cMpc LAE filament. This structure connects two protocluster nodes, but both show an approximately 25 cMpc offset to the quasar density peak.
The inspection of LAE properties, including luminosity, potential AGN activity, and extended \lya~emission, reveals significant differences between the two nodes.

\item A 3D IGM \textsc{Hi} tomography map is reconstructed from SDSS/eBOSS, which reveals a dual ionization state along the LAE filament: nearly fully ionized IGM around Node-A at NW, contrasting with a transition to more neutral states towards the Node-B at SE. 
This variance corroborates a similarly distinctive characteristic found by LAE properties between the two nodes.

\item The analysis with the 3D IGM \textsc{Hi} tomography map indicates that quasars are preferentially located near the boundaries between ionized and neutral IGM regions. 
This spatial distribution, alongside a significant positive correlation between quasar bolometric luminosity ($L_{\rm bol}$) and local \textsc{Hi} transmission fluctuation ($\delta_{F}$), underscores a potential physical association between clustered 
quasar activity and large-scale IGM ionization.

\item The offset between the peaks of quasars and LAEs challenges conventional theories of quasar activation solely through major mergers. It either suggests that a significant amount of other galaxies are missing or killed in the quasar proximity, or it hints at complex triggering mechanisms, potentially including intense environmental interactions reminiscent of cosmic structures like the Bullet Cluster in the local universe.
\end{itemize}

We have discussed some potential scenarios with mixtures of {\sc Hi} overdensities and quasar photoionization to explain a part of the observed results, but no conclusive theory can fully resolve the emergence of \textit{Cosmic Himalayas}. 
The presented findings, supported by rigorous data analyses, not only shed light on the relationships between quasars, galaxies, and the IGM in an extreme cosmic structure, the \textit{Cosmic Himalayas}, but also pave the way for future multi-wavelength, high-resolution observational campaigns aimed at unraveling the intricate evolution of such remarkable cosmic phenomena.

\vspace{20pt}

We thank Yechi Zhang, Yuki Isobe, Yi Xu, Hiroya Umeda, Khee-Gan Lee, Antonio Pensabene, Titouan Lazeyras, Marta Galbiati, Matteo Viel, Xianzhong Zheng, Veronica Strazzullo, Maurilio Pannella, Alex Saro, Stefano Borgani, Laura Pentericci, Mark Dickinson, Lorenzo Napolitano, Paolo Cassata, Giulia Rodighiero, Olga Cucciati, Roberto Decarli, Akio K. Inoue, Yuqing Lou, Guochao Sun for helpful discussions.
This work was supported by JSPS KAKENHI Grant Number JP20H00180 (PI: M. Ouchi) and JP24K17084 (PI: Y. Liang). 
Y.L. acknowledges the travel support from the program International Leading Research “Comprehensive understanding of the formation history of structures in the Universe”.

Funding for the Sloan Digital Sky Survey V has been provided by the Alfred P. Sloan Foundation, the Heising-Simons Foundation, the National Science Foundation, and the Participating Institutions. SDSS acknowledges support and resources from the Center for High-Performance Computing at the University of Utah. The SDSS web site is \url{www.sdss.org}.

SDSS is managed by the Astrophysical Research Consortium for the Participating Institutions of the SDSS Collaboration, including the Carnegie Institution for Science, Chilean National Time Allocation Committee (CNTAC) ratified researchers, the Gotham Participation Group, Harvard University, Heidelberg University, The Johns Hopkins University, L’Ecole polytechnique f\'{e}d\'{e}rale de Lausanne (EPFL), Leibniz-Institut f\"{u}r Astrophysik Potsdam (AIP), Max-Planck-Institut f\"{u}r Astronomie (MPIA Heidelberg), Max-Planck-Institut f\"{u}r Extraterrestrische Physik (MPE), Nanjing University, National Astronomical Observatories of China (NAOC), New Mexico State University, The Ohio State University, Pennsylvania State University, Smithsonian Astrophysical Observatory, Space Telescope Science Institute (STScI), the Stellar Astrophysics Participation Group, Universidad Nacional Aut\'{o}noma de M\'{e}xico, University of Arizona, University of Colorado Boulder, University of Illinois at Urbana-Champaign, University of Toronto, University of Utah, University of Virginia, Yale University, and Yunnan University.

This research is based in part on data collected at the Subaru Telescope, which is operated by the National Astronomical Observatory of Japan. We are honored and grateful for the opportunity of observing the Universe from Maunakea, which has the cultural, historical, and natural significance in Hawaii.
The Hyper Suprime-Cam (HSC) collaboration includes the astronomical communities of Japan and Taiwan, and Princeton University. The HSC instrumentation and software were developed by the National
Astronomical Observatory of Japan (NAOJ), the Kavli
Institute for the Physics and Mathematics of the Universe (Kavli IPMU), the University of Tokyo, the High
Energy Accelerator Research Organization (KEK), the
Academia Sinica Institute for Astronomy and Astrophysics in Taiwan (ASIAA), and Princeton University.
This paper makes use of software developed for the Large Synoptic Survey Telescope. We thank the LSST Project for making their code available as free software at \url{http://dm.lsst.org}.

Data analysis was in part carried out on the Multi-wavelength Data Analysis System operated by the Astronomy Data Center (ADC), National Astronomical Observatory of Japan.

{\it Facility:} Sloan, Subaru (HSC), GALEX

% \end{acknowledgments}


\bibliography{bib_liang24}{}
\bibliographystyle{aasjournal}



\end{document}